\documentclass[
   ,final            
  ,numberedheadings 
  ]
 {aipproc}

\layoutstyle{6x9}

\usepackage{amsfonts}
\usepackage{amssymb}
\usepackage{mathrsfs}

\newcommand{\be}{\begin{equation}}
\newcommand{\ee}{\end{equation}}
\newcommand{\ba}{\begin{array}}
\newcommand{\ea}{\end{array}}
\newcommand{\bea}{\begin{eqnarray}}
\newcommand{\eea}{\end{eqnarray}}
\newcommand{\Lfr}[1]{\mbox{\Lfrak #1}}  
\newcommand{\bal}{\begin{align}}
\newcommand{\eal}{\end{align}}
\newcommand{\trip}{|\! |\! |}
\font\Lfrak=eufm10 at 12 pt

\usepackage{color}

\def\Hb{\mathbb{H}}
\def\Rb{\mathbb{R}}

\begin{document}

\title{The fundamental role of symmetry in nuclear models\\ \vspace{0.2cm}
{\normalsize Lecture notes for the International Scientific Meeting on Nuclear Physics\\ 
 at La R\'abida, Huelva (Spain), September 9--13, 2012. } }


\author{D. J. Rowe}{
  address={Department of Physics, University of Toronto, Toronto, 
  Ontario M5S 1A7, Canada}   }

\begin{abstract}
The purpose of these lectures is to illustrate how symmetry and pattern recognition play essential roles in the progression from experimental observation to an understanding of nuclear phenomena in terms of interacting neutrons and protons.  
We do not discuss weak interactions nor relativistic and sub-nucleon degrees of freedom.
The explicit use of symmetry and the power of algebraic methods, in combination with analytical and geometrical methods are illustrated by their use in deriving a shell-model description of nuclear rotational dynamics and the structure of deformed nuclei. 
\end{abstract}

\maketitle

\section{Introduction}

It has become common practice in nuclear physics to make a distinction between algebraic and geometric models.  This is convenient for some purposes but, in general, it is misleading because almost all models in quantum mechanics have expressions in algebraic terms.  In fact, many-particle quantum  mechanics is fundamentally a unitary representation of the Lie algebra of one-body operators and its observables are polynomials in the elements of this Lie algebra.

In proceeding towards a microscopic understanding of some nuclear phenomena, it is profitable to follow a sequence of steps along the following lines:
\begin{enumerate}
\item[(i)]  Observe the phenomena in many situations until its pattern becomes evident.
\item[(ii)]  Develop a phenomenological model to explain the observations and suggest new observations to ascertain the consistency of the  model.
\item[(iii)] Repeat steps (i) and (ii) to refine the model, assess its reliability and domain of validity, and identify its algebraic structure.  
\item[(iv)]  If a model provides a successful  understanding of the phenomena, the next step is to understand the model in terms of interacting nucleons.  This is achieved if one is able to express the observables of the model in terms of nucleon coordinates and momenta.  In the process, it is likely that the model will only be realisable in some limit or as an approximation to a more complex microscopic theory. It will then be of fundamental importance to determine if the limitations are consistent with the experimental observations.
As the following examples will illustrate, this exercise can also lead to improved phenomenological models with clearer microscopic foundations.
\item[(v)]  Having found a microscopically successful  model, whose representations are capable of explaining a range of physical observations, the next goal is to make use of its algebraic structure to identify an associated shell-model coupling scheme that will be appropriate for a shell-model description of the phenomena with realistic nucleon-nucleon interactions.
\end{enumerate}

The remarkable fact is that by such sequences of steps, one has  obtained shell model coupling schemes appropriate for some of the dominant characteristics of nuclei. For example,  the basic harmonic-oscillator shell-model,  proposed to explain the strong binding energy of so-called magic nuclei in terms of closed-shell structures, has provided the independent-particle basis for the standard shell model given by nucleons in a mean-field with spin-orbit interactions.  Pairing models of singly closed-shell nuclei lead to the standard $jj$-coupling scheme of Flowers \cite{Flowers52}; and nuclear rotational models lead to the Elliott SU(3) coupling scheme \cite{Elliott58ab} for light nuclei.

\begin{figure}
  \includegraphics[height=.4\textheight]{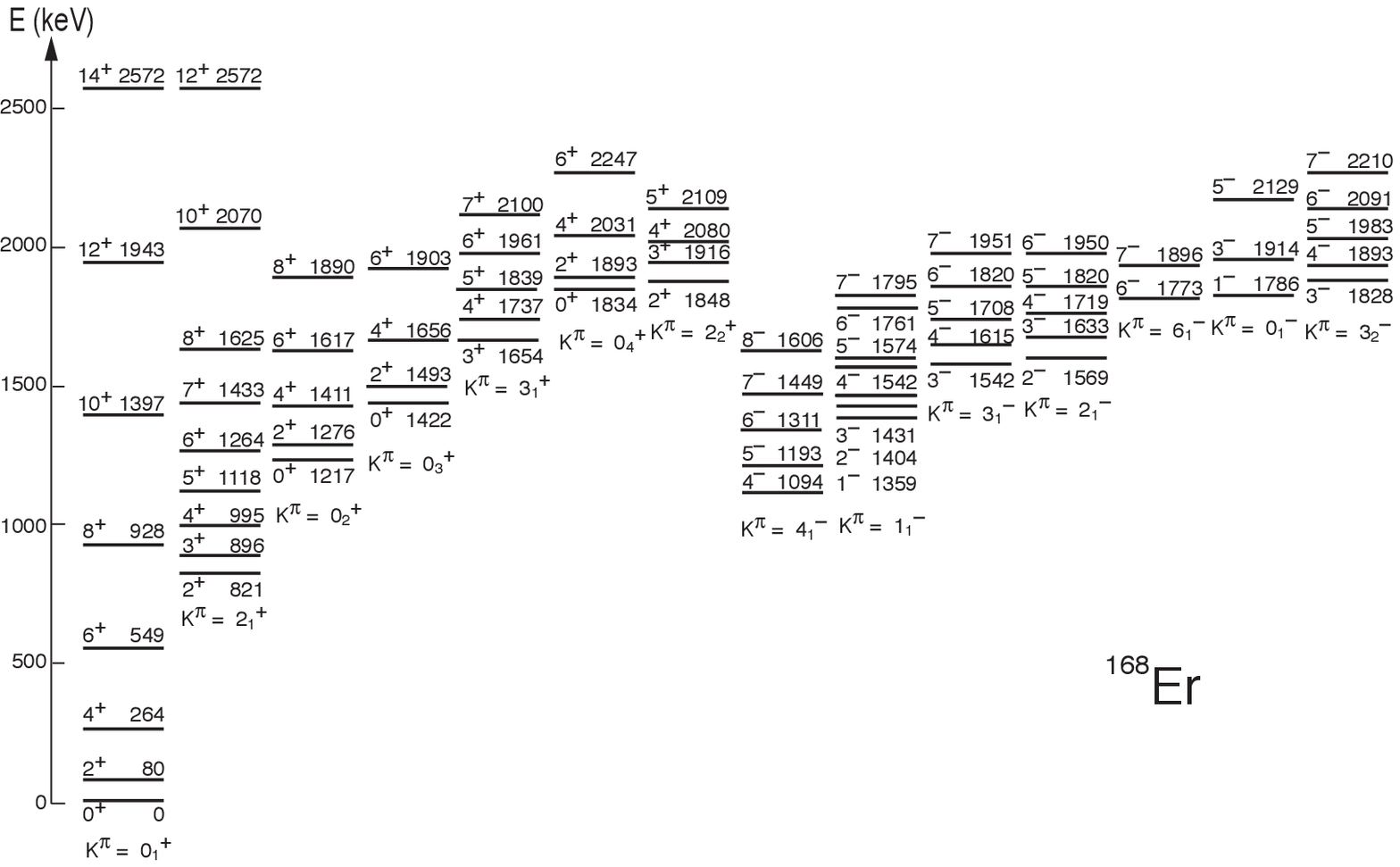}
  \caption{\label{fig:168Er} The low-lying states of $^{168}$Er arranged into rotational bands.}
\end{figure}

In the following we start from the observation of rotational bands organised as shown, for example, in Fig.\ \ref{fig:168Er} and the interpretations of such bands in the Bohr model and proceed to their implications for a shell-model theory of strongly deformed nuclei.
However, before embarking on this process, we first recall some of the primary symmetry concepts that will be used.

\section{Symmetry groups, dynamical groups and spectrum generating algebras} 

\noindent{\bf Definition}:
{A symmetry group of a system is a group of transformations of the system that leave its Hamiltonian invariant.}  \\

There are several variations  in the literature on the definition of a dynamical group \cite{BohmNB88}.  The following definition is most useful for present purposes.\\

\noindent{\bf Definition}:
{A dynamical group for a Hamiltonian  $\hat H$ is a Lie group of  unitary transformations of the  Hilbert space $\Hb$  of a system with Hamiltonian 
 $\hat H$ such that the subspaces of $\Hb$ that are invariant under the transformations of the dynamical group are spanned by eigenstates of 
 $\hat H$.}\\

 Consider, for example, a particle in ordinary three space moving in a central force potential $V(r)$ such that the Hamiltonian is  SO(3) invariant.
 Basis wave functions for the particle's Hilbert space are  given by products of radial wave functions and spherical harmonics
 \be \psi_{nlm}(r,\theta,\varphi) = R_{nl}(r) Y_{lm} (\theta\varphi),\ee
where $l$ and $m$ are angular-momentum quantum numbers and $n$ is a radial quantum number.
The Hamiltonian matrix in this basis is  given by
\be H^l_{nm,n'm'} = \langle \psi_{nlm} |\hat H|\psi_{n'lm'}\rangle \ee
which, because SO(3) is a symmetry group, has $l$ as a good quantum number.
In addition, because $\hat H$ is SO(3) invariant, the Hamiltonian matrix elements
are diagonal in $m$ and independent of its value, i.e.,
\be H^l_{nm,n'm'} = \delta_{m,m'} H^l_{n,n'} \ee
and $m$ is also a good quantum number.
Thus, to obtain the energy spectrum for the particle, it remains only to diagonalise each $H^l$ with respect to the radial quantum number $n$ to obtain its eigenfunctions
\be \Psi_{\alpha lm} = \sum_n C_{\alpha n} \psi_{nlm}.\ee
This process is greatly simplified by the fact that the central force problem has a dynamical group SU(1,1) for which the set of states 
$\{ \psi_{nlm} ; n=0, 1,2, \dots\}$ for each pair of  $lm$ values carries an irreducible representation.\\

\noindent{\bf Definition}:
A spectrum generating algebra (SGA) for a Hamiltonian is a Lie algebra such that the Hamiltonian can be expressed (most usefully as polynomials) in terms of its elements.\\

An SGA $\Lfr{g}$ is most useful because its representation theory enables one to calculate the matrix elements of any $X\in \Lfr{g}$ and hence the matrix elements of the Hamiltonian.  A Hamiltonian may have many SGA's.  A particularly useful choice is one for which the important observables of the system have simple expressions in terms of its Lie algebra $\Lfr{g}$, e.g.,  as linear or quadratic polynomials of Lie algebra elements.

An SGA for the above central-force problem is the Lie algebra su(1,1) whose infinitesimal generators include the square of the radial coordinate, $r^2$, of the particle and the square of it momentum, $p^2$.

A system may  have many symmetry groups and many dynamical groups.  
For example, the group generated by an SGA for a Hamiltonian is a dynamical group for that Hamiltonian.

\section{The Bohr collective model}

The Bohr model is a liquid-drop model with quadrupole shape coordinates  defined for its surface radius by 
$$ R(\theta,\varphi) = 
R_0 \Big[ 1 + \sum_\nu \alpha_\nu Y^*_{2\nu}(\theta,\varphi) + \dots \Big] .$$
It is quantised by the introduction of quadrupole shape observables 
$\{ \hat \alpha_\nu\}$ and canonical momenta $\{\hat \pi^\nu\}$ that satisfy the Heisenberg commutation relations
$$ [\hat \alpha_\mu, \hat \pi^\nu] = {\rm i} \hbar \delta_\mu^\nu $$
and act as operators on a Hilbert space $\Hb$ of square-integrable wave functions of the shape variables, according to the equations
$$ \hat \alpha_\nu \psi(\alpha) = \alpha_\nu \psi(\alpha) , \quad
\hat\pi^\nu = -{\rm i}\hbar \frac{\partial}{\partial \alpha_\nu}\psi(\alpha),
\quad \nu = 0,\, \pm 1,\, \pm 2 . $$
A Hamiltonian for the model is then an SO(3)-invariant operator of the form
$$ \hat H = \frac{1}{B}  \hat  \pi \cdot \hat \pi + V(\alpha) ,$$
where $V(\alpha)$ is a rotationally-invariant potential energy (a function of the coordinates) and $B$ is a  mass parameter.

\begin{figure}[ht]
\includegraphics[width=30pc]{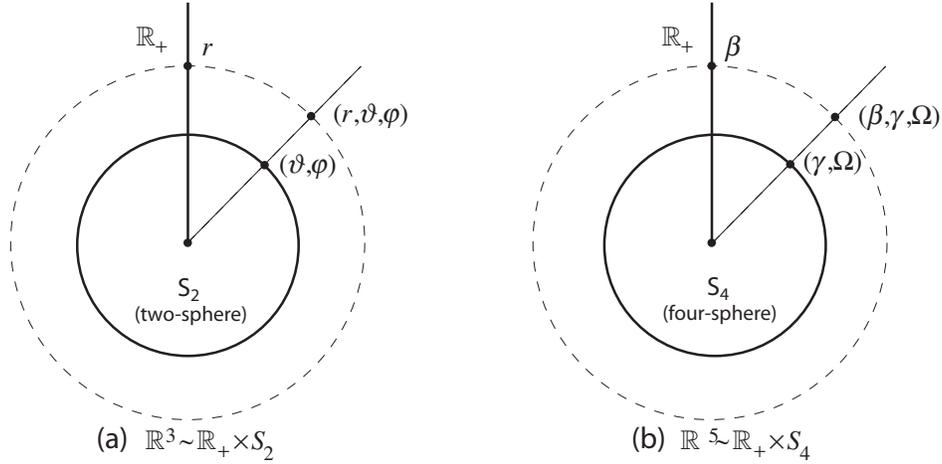}
\caption{Spherical polar coordinates (a) for a  three-dimensional and (b) for a five-dimensional Euclidean space.
\label{fig:BohrGeom}}
\end{figure}

The model has  geometric and algebraic structures that make it  easy to use, but which have only recently been exploited in what is known as the Algebraic Collective Model (ACM) \cite{RoweT05,CaprioRW09,RoweWC09,WelshR13}.
The coordinate space of this model is the real five-dimensional 
Euclidean space $\Rb^5$.  This space has much in common with the more familiar three-dimensional space $\Rb^3$ and similarly, for a rotationally-invariant Hamiltonian, it is most naturally assigned spherical polar coordinates.
Apart from the origin (which is a point of measure zero), the Euclidean space
$\Rb^3$ is a tensor product of a radial line and a two-sphere of unit radius as illustrated in Fig.\ \ref{fig:BohrGeom}(a).
Similarly, to within a point of measure zero, the Euclidean space
$\Rb^5$ is a tensor product of a radial line and a four-sphere of unit radius.
Thus, whereas $\Rb^3$ is  assigned spherical polar coordinates
$(r, \theta,\varphi)$, where $r$ is a radial coordinate and $(\theta,\varphi)$ are coordinates for a point on the two-sphere, the space $\Rb^5$ is assigned spherical polar coordinates $(\beta,\gamma,\Omega)$, where now $\beta$ is a radial coordinate, defined by $\beta^2 = \alpha\cdot \alpha$, which together with an angle coordinate $\gamma$ determine the triaxial shape of a rotor;
$\Omega$ denotes the orientation angles of this rotor. Together, the angle coordinates $(\gamma,\Omega)$ complete a system of spherical coordinates for the four-sphere.
Thus, whereas the Hilbert space for $\Rb^3$ is the tensor product 
\be {\cal L}^2(\Rb^3) = {\cal L}^2(\Rb_+)\times {\cal L}^2(S_2) \ee
and is spanned by functions  
$\{\psi_{nlm}(r,\theta,\varphi)= R_{ln}(r) Y_{lm} (\theta,\varphi)\}$, where 
$\{R_{ln}(r)\}$ is a basis for a unitary irrep (irreducible representation) of the group SU(1,1) and 
$\{Y_{lm} (\theta,\varphi)\}$is a basis of spherical harmonics for the two-sphere, the Hilbert space for $\Rb^5$ is the tensor product 
\be {\cal L}^2(\Rb^5) = {\cal L}^2(\Rb_+)\times {\cal L}^2(S_4) \ee
and is spanned by functions  
$\{ {\cal R}_{vn}(\beta) {\cal Y}_{v\alpha LM} (\gamma,\Omega)\}$, where 
$\{{\cal R}_{vn}(\beta)\}$ is again a basis for a unitary irrep  of the group SU(1,1) and  $\{\mathcal{Y}_{v\alpha LM} (\gamma,\Omega)\}$ is a basis of SO(5) spherical harmonics for the four-sphere.

Matrix elements of  operators of interest on the Hilbert space
${\cal L}^2(\Rb^3)$ can be calculated algebraically.  This is because  explicit expressions are known \cite{Rowe05} for the matrix elements of $r, 1/r, r^2, 1/r^2, \nabla^2$, for  SO(3) spherical harmonics, for Clebsch-Gordan coefficients, and for the SO(3)-reduced matrix elements
\be \langle Y_{l_3} \| \hat Y_{l_2} \| Y_{l_1}\rangle
= \sqrt{\frac{(2l_2+1)(2l_1+1)}{4\pi}} \, (l_1 0 \, l_2 0 |l_3 0) .\ee

Similar expressions are now known for the Hilbert space ${\cal L}^2(\Rb^5)$ of the Bohr model.
By simple algebraic methods  \cite{Rowe05}, we obtain explicit expressions for the  matrix elements of $\beta, 1/\beta, \beta^2, 1/\beta^2$ and $\nabla^2$. From the basic SO(5) spherical harmonics
\bea
&& {\cal Y}_{1122} (\gamma,\Omega) \propto \cos\gamma {\cal D}^2_{02} + 
{\textstyle\sqrt{\frac12}} \sin\gamma 
\left( {\cal D}^2_{22}(\Omega) + {\cal D}^2_{-2,2}(\Omega) \right), \\
&& {\cal Y}_{2122} (\gamma,\Omega) \propto \cos 2\gamma {\cal D}^2_{02} - 
{\textstyle\sqrt{\frac12}} \sin2\gamma
\left( {\cal D}^2_{22}(\Omega) + {\cal D}^2_{-2,2}(\Omega) \right),\\
&& {\cal Y}_{3100} (\gamma,\Omega) \propto \cos 3\gamma,\\
&& {\cal Y}_{3133} (\gamma,\Omega) \propto {\textstyle\sqrt{\frac12}}
\sin 3\gamma \left( {\cal D}^3_{23}(\Omega) - {\cal D}^2_{-2,3}(\Omega) \right) ,
\eea
we can generate the complete set of SO(5) spherical harmonics 
 and, from them, derive all  required SO(5) Clebsch-Gordan coefficients 
 \cite{RoweTR04,CaprioRW09}.  We also obtain \cite{CaprioRW09} the SO(5)-reduced matrix elements
\bea
\langle {\mathcal{Y}}_{v_3} \trip \hat{\mathcal{Y}}_{v_2}
 \trip {\mathcal{Y}}_{v_1}\rangle  &=& \frac{1}{4\pi}  
\frac{(\frac\sigma2+1)!}
{(\frac\sigma2-v_1)!(\frac\sigma2-v_2)!(\frac\sigma2-v_3)!} 
 \sqrt{\frac{(2v_1+3)(2v_2+3)}{(v_3+2)(v_3+1)}}\,  \nonumber\\
&& \times \sqrt{
\frac{(\sigma+4)(\sigma-2v_1+1)!(\sigma-2v_2+1)!(\sigma-2v_3+1)!}
     {(\sigma+3)!}}, \qquad \label{eq:3.claim5}
\eea 
where $\sigma=v_1+v_2 +v_3$. 
 Thus, Bohr model calculations are carried out quickly and easily.

\begin{figure}
  \includegraphics[height=.43\textheight]{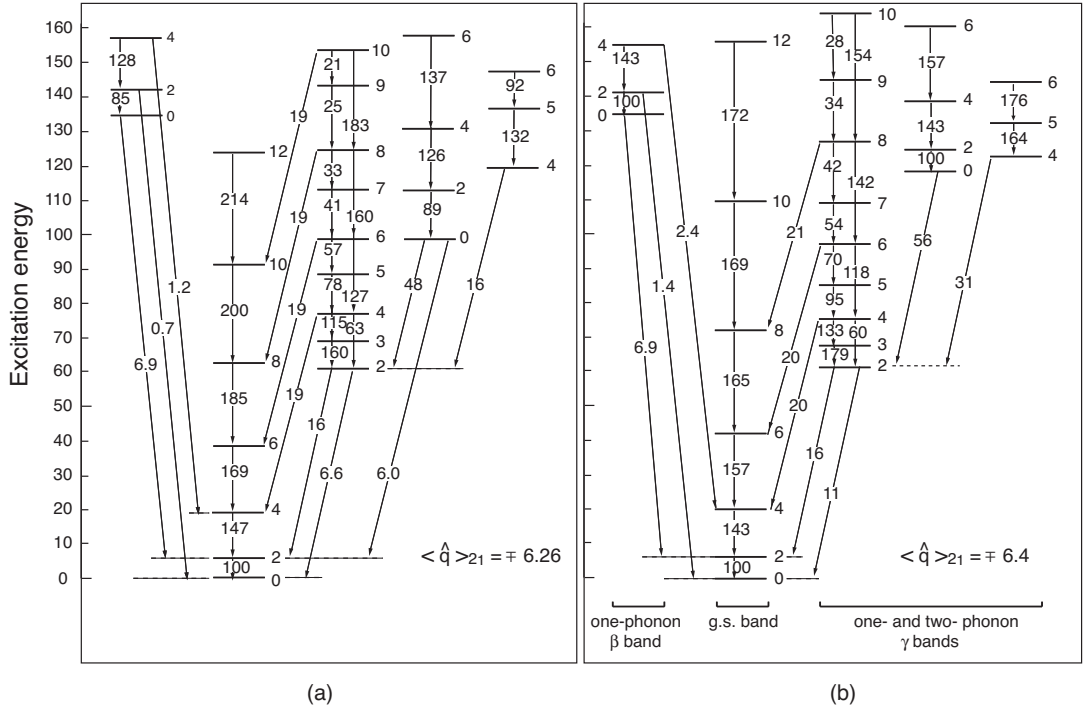}
  	\caption{(a) The low-energy spectrum of the Hamiltonian 
    $\hat H(B,\lambda,\chi,\kappa=0)$ of Equation (\protect\ref{eq:ACMHgen}) 
    for  $B=20$, $\lambda = 1.5$,  and $\chi = \pm 2.0$.     
     Reduced E2 transition rates are shown in units for
    which $B({\rm E2}; 2_1\to 0_1) =100$. Energy levels are given in units 
    such that the lowest $L=2$ state has energy $E_{2_1}= 6$. 
    (b)  The corresponding results for the adiabatic Bohr model in the
	axially symmetric limit showing a ground-state band,
       a one-phonon beta vibrational band, and one- and two-phonon
       gamma vibrational bands. (Figure from  \cite{RoweWC09}) }
	\label{fig:BM1}
\end{figure}

Particularly informative are Bohr Hamiltonians are of the form
\be \hat H(B,\lambda,\chi,\kappa) =   -\frac{\nabla^2}{2B} + 
V_\lambda(\beta,\gamma) \ee
with
\be V_\lambda(\beta,\gamma)=
\frac12  B\big[(1-2\lambda)\beta^2 + \lambda \beta^4\big] 
- \chi\beta\cos 3\gamma +  \kappa \cos^2 3\gamma  .  \label{eq:ACMHgen}
\ee
For example,  Fig.\ \ref{fig:BM1}(a) shows the low energy-level spectrum and E2 transition  rates for such a Hamiltonian with $\kappa = 0$.
The physical content of this spectrum is exposed by comparing it with results for the simpler Hamiltonian with potential
\be V(\beta,\gamma) \approx V(\beta_0,\gamma_0) + 
\frac12 B \omega_\beta^2
(\beta-\beta_0)^2 + \frac12 B\beta_0^2 \omega_\gamma^2
(\gamma-\gamma_0)^2 , \label{eq:Vadiabatic}
\ee
where $\beta_0=0$ and $\gamma_0=0$ are the  values of $\beta$ and $\gamma$ at the potential minimum.  With the additional assumption that the rotational motions of the model are slow relative to its $\beta$ and $\gamma$ vibrational
modes there is an adiabatic decoupling of the rotational and vibrational degrees of freedom and the spectrum that emerges becomes as shown in 
Fig.\ \ref{fig:BM1}(b).  Thus, the relatively small differences between Figs.\ 
\ref{fig:BM1}(a) and \ref{fig:BM1}(b) can be attributed to SO(5) centrifugal coupling effects between the rotational and vibrational degrees of freedom that are included in Fig.\ 
\ref{fig:BM1}(a) but not in \ref{fig:BM1}(b).

The Bohr model and its many generalisations by Bohr, Mottelson, and numerous colleagues \cite{BohrM75} has been enormously successful in the interpretation of a huge body of experimental data on collective motions.
A particularly significant generalisation was the extension of the Bohr model to describe the dynamics of a nucleon coupled to an even-mass rotational nucleus \cite{Nilsson55}.
This and other generalisations reveal how nucleons in a deformed mean field can form a rotor with many intrinsic degrees of freedom that can rotate adiabatically without undue disturbance of its intrinsic structure.
Such generalisations are known as unified models.

\section{Progression to a microscopic collective model}
 
 In view of the successes of the Bohr and unified models, a great deal of effort has been expended in seeking to express  collective dynamics in terms of microscopic nucleon coordinates.  
In retrospect, this is simple if one starts by replacing  the surface shape coordinates $\{ \alpha_\nu\}$ of the Bohr model by microscopic Cartesian monopole/quadrupole moments
\be Q_{ij} = \sum_{n=1}^A x_{ni} x_{nj} , \quad i,j = 1,2,3, \;\; n=1,\dots, A. \ee
Time derivatives and corresponding momentum observables are then given by
\be \dot Q_{ij} = \frac{dQ_{ij}}{dt} =\sum_n (\dot x_{ni} x_{nj} + x_{ni} \dot x_{nj}) , \quad P_{ij} = M \dot Q_{ij} =\sum_n ( p_{ni} x_{nj} + x_{ni} p_{nj}), \label{eq:P&Q}\ee
where $M$ is the nucleon mass.
These  shape and momentum  observables
are now microscopic and close on a Lie algebra with   commutation relations,
\be [\hat  Q_{ij}, \hat P_{kl}] =  {\rm i}\hbar \big[ \delta_{il}\hat Q_{jk} +  \delta_{ik}  \hat Q_{jl} + \delta_{jl} \hat  Q_{ik} + \delta_{jk}  \hat Q_{il}  ]. \label{eq:QPalg} \ee

This Lie algebra differs from that of the Bohr model.  However, the subset of  quadrupole moments defined in a spherical tensor basis by
\bea && \hat q_0 = {\textstyle \frac{1}{\sqrt{8}}}\,\varepsilon 
(2\hat Q_{11} - \hat Q_{22} -\hat  Q_{33}),\nonumber\\
&& \hat q_{\pm 1} = {\textstyle \sqrt{\frac34}}\,\varepsilon 
(\hat Q_{12} \pm {\rm i} \hat Q_{12}),\\
&& \hat q_{\pm 2} =  {\textstyle \frac{\sqrt{3}}{4}}\,\varepsilon 
(\hat Q_{22}- \hat Q_{33} \pm2{\rm i} \hat Q_{23}),\nonumber
\eea
have commutation relations with the corresponding quadrupole  momenta that contract to those of the Bohr model when restricted to states of the Hilbert space of large monopole and relatively small quadrupole moments.  For example, with the monopole moment defined by 
\be \hat{\Lfr{M}}_0 = \hat Q_{11} + \hat Q_{22} + \hat Q_{33}, \ee
and $\varepsilon^2 = 1/\langle \hat{\Lfr{M}}_0 \rangle$, it follows that
\be [\hat q_\mu, \hat p^\nu] = {\rm i} \hbar \delta_\mu^\nu + 0(\varepsilon^2)
\to  {\rm i} \hbar \delta_\mu^\nu ,   \quad{{\rm as\;} \varepsilon\to 0} .
\ee
We refer the model with this SGA as the microscopic Bohr Model.

In spite of this  result, this model fails to describe the low-energy rotational bands for which the Bohr model was introduced. 
This is because its momentum operators are the infinitesimal generators of irrotational-flow rotations for which the moments of inertia are much smaller than those needed to describe observed rotational bands.  
However,  this limitation  is overcome by enlarging the Lie algebra, spanned by
$\{ \hat Q_{ij}, \hat P_{ij}\}$, to include the angular momentum operators
\be \hat L_k = \hbar \hat L_{ij} 
= \sum_n \big(\hat x_{ni}\hat p_{nj}  -\hat x_{nj}\hat p_{ni}\big) . \ee
The Lie algebra obtained is then isomorphic to a known CM(3) Lie algebra introduced by Weaver \emph{et al.} \cite{WeaverBC73}.

\begin{figure}[ht]
\includegraphics[width=18pc]{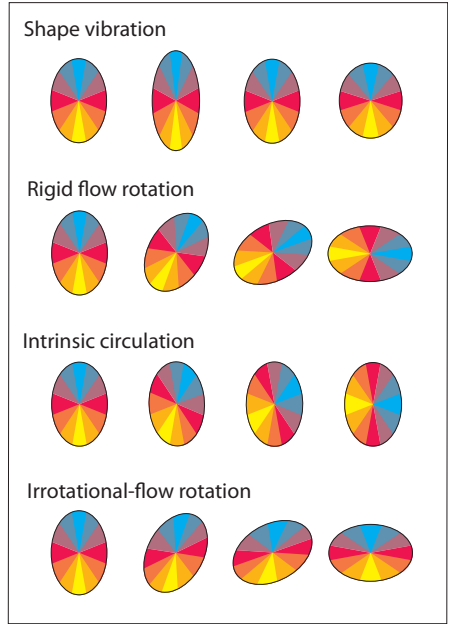}
\caption{Collective flows generated by CM(3) transformations
\label{fig:GLFlows}}
\end{figure}

The CM(3) model is  richer than the Bohr model.  Its six independent
 $\{ \hat Q_{ij}\}$ observables characterize the size and quadrupole shape of a nucleus and, among its 9 momenta  $\{\hat L_{ij}, \hat P_{ij}\}$, the angular momenta $\hat L_{ij}$ are infinitesimal generators of rigid-body rotations, the three components of $\hat P_{ij}$ with $i\not= j$ are infinitesimal generators of irrotational-flow rotations, and the three components $\hat P_{ii}$ are infinitesimal generators of diagonal shape scale transformations \cite{RoweR80}.  Combinations of these momentum operators also generate intrinsic current circulations as illustrated in the third row of Fig.\ 
 \ref{fig:GLFlows}.
%

The unitary irreps of CM(3) were determined by Rosensteel \cite{RosensteelR76}
with the intention of developing a microscopic theory of collective motion. However, two obstacles were encountered: one was that, while the many-nucleon kinetic energy has a simple known expression, 
$T= \frac{1}{2m}\sum_{ni} \hat p_{ni}^2$, its CM(3) collective model component (derived for example in  \cite{RoweR79}) proved to be excessively complicated; another was that the CM(3) representations did not relate in a natural way to the nuclear shell model.
The way around both obstacles was simply to augment the CM(3) algebra further so that it included the many-nucleon kinetic energy.
The result was the much more powerful symplectic model.

\section{The symplectic model}

The symplectic model  \cite{RosensteelR77a,RosensteelR80} is an algebraic model whose dynamical group Sp(3,\,R) comprises all $n$-independent linear canonical transformations of the single-particle phase-space observables  
 $$ \hat x_{ni} \to \sum_j \big(a_{ij}\hat  x_{nj} + b_{ij} \hat p_{nj}\big) , \quad
 \hat p_{ni} \to \sum_j \big(c_{ij} \hat x_{nj} + d_{ij} \hat p_{nj}\big) ,$$
  that preserve the commutation relations 
 $[\hat x_{ni}, \hat p_{mj}] = {\rm i} \hbar \delta_{m,n}\delta_{i,j}$.
Its spectrum generating algebra is spanned by the  elements of the CM(3) Lie algebra
\be \hat Q_{ij} = \sum_{n=1}^A \hat x_{ni} \hat x_{nj} , \quad 
\hat P_{ij} = \sum_n ( \hat p_{ni} \hat x_{nj} + \hat x_{ni} \hat p_{nj}), \quad
\hbar \hat L_{ij} = \sum_n \big(\hat x_{ni}\hat p_{nj}  -\hat x_{nj}\hat p_{ni}\big),\ee
plus the  bilinear  momentum operators
\be \hat K_{ij}= \sum_n  \hat p_{ni}  \hat p_{nj} .\ee

Adding the  $\hat K_{ij}$ operators to the CM(3) algebra is invaluable for many reasons:

\begin{itemize}

\item[(i)]  The representations of Sp(3,\,R) are easier to handle than those
 of CM(3). 

\item[(ii)]  The many-nucleon kinetic energy and harmonic-oscillator Hamiltonians 
\be  \sum_{n=1}^A \sum_{i=1}^3 \frac{1}{2M} \hat {p}_{ni}^2 , \quad  
\hat H^{(A)}_{\rm DHO} = \sum_{n=1}^A
\left[ \frac{1}{2M} \hat {\bf p}_n^2 
+\frac12 M (\omega_1^2 \hat x^2_{n1} + \omega_2^2 \hat x^2_{n2} +
\omega_3^2 \hat x^2_{n3})\right] ,
\ee
where $M$ is the nucleon mass, are elements of the Sp(3,\,R) Lie algebra for all values of $\omega_1, \omega_2,  \omega_3$.
As a result, the symplectic model is compatible with the harmonic oscillator shell model and with unified collective models such as the Nilsson model.

\item[(iii)]  Sp(3,\,R) representation theory makes it possible to diagonalize a collective model Hamiltonian
\be \hat H = \frac{1}{2M} \sum_i \hat K_{ii} + V(Q) ,\ee
in a basis of  shell-model states.\\

\item[(iv)]  The  group, Sp(3,\,R), contains  important subgroup chains
\bea
&& {\rm Sp(3,\,R)} \supset {\rm CM(3)} \supset {\rm ROT}(3) \supset {\rm SO(3)}  ,
\nonumber \\
&&  {\rm Sp(3,\,R)} \supset {\rm U(3)} \supset {\rm SU}(3) \supset {\rm SO(3)}  ,
\nonumber 
\eea
where ROT(3) is a dynamical group for a rigid-rotor model and SU(3) is the dynamical group of Elliott's SU(3) model.
The first chain reflects its  content in terms of the dynamical flows of a quantum fluid.  The latter  chain defines an optimal shell-model coupling scheme for a shell-model description of collective states in nuclei. 

\end{itemize}

These properties are   all you could wish for from a microscopic collective model.

\subsection{The Sp(3,\,R) algebra in a U(3) basis}

The  U(3) $\subset$ Sp(3,\,R) subalgebra   consists of all elements of the  
Sp$(3,\Rb)$ Lie algebra  that commute with the spherical harmonic-oscillator Hamiltonian
\be\hat H^{(A)}_{\rm SHO} =\sum_{n=1}^A \left[ \frac{1}{2M} \hat {\bf p}_n^2 
+\frac12 M\omega^2 ( \hat x^2_{n1}+ \hat x^2_{n2}+ \hat x^2_{n3})\right] ,
\label{eq:Hsho}
\ee
where $\omega$ is fixed to get the mean-square radius of a nucleus at its observed value.  

To identify this subalgebra, express the $\hat x_{ni}$ and $\hat p_{ni}$ observables in terms of harmonic oscillator raising and lowering operators 
\be \hat x_{ni} = \frac{1}{\sqrt{2}\,a} (c^\dagger_{ni} + c_{ni} ) , \quad
\hat p_{ni} = {\rm i} \hbar\frac{a}{\sqrt{2}} (c^\dagger_{ni} - c_{ni} ) ,
\label{eq:xpccdag} \ee
where $a  = \sqrt{M\omega/\hbar}$ is the harmonic oscillator unit of inverse length.
These operators satisfy the commutation relations
\be  [c_{ni}, c^\dagger_{mj}] = \delta_{m,n} \delta_{i,j}, \quad
  [\hat H^{(A)}_{\rm SHO}, c^\dagger_{ni}]= \hbar\omega c^\dagger_{ni}, \quad
  [\hat H^{(A)}_{\rm SHO}, c_{ni}]= -\hbar\omega c_{ni}.
\ee  
Expansion of the Sp$(3,R)$ operators in terms of them gives
\bea &&a^2 \hat Q_{ij} = \hat{\mathcal{Q}}_{ij}  + \hat A_{ij} + \hat B_{ij},
\label{eq:Qexpand}\\
&& \hat P_{ij} = 2{\rm i}\hbar (\hat A_{ij} - \hat B_{ij}) ,\nonumber\\
&&\frac{1}{a^2} \hat K_{ij} = \hbar^2 ( \hat{\mathcal{Q}}_{ij} 
- \hat A_{ij} - \hat B_{ij} ) , \nonumber\\
&&\hat L_{ij}  = -{\rm i} ( \hat C_{ij} - \hat C_{ji} ) , \nonumber
\eea
where
\be \begin{array}{l}
 \hat A_{ij} = \hat A_{ji} = \textstyle\frac12\sum_n c^\dagger_{ni} c^\dagger_{nj} ,\quad
\hat B_{ij} = \hat B_{ji} =\textstyle\frac12\sum_n c_{ni} c_{nj} , \\
\hat C_{ij} =  \sum_n \big( c^\dagger_{ni} c_{nj} + \textstyle\frac12\big), \quad
\hat{\mathcal{Q}}_{ij} = \textstyle\frac12\big( \hat C_{ij} + \hat C_{ji} \big) ,
\end{array}
\ee
satisfy the commutation relations
\be \begin{array}{l}
 [\hat H^{(A)}_{\rm SHO}, \hat{\mathcal{Q}}_{ij}]=0, \quad
[\hat H^{(A)}_{\rm SHO}, \hat{L}_{ij}]=0,  \cr
[\hat H^{(A)}_{\rm SHO},\hat A_{ij}]=2\hbar\omega \hat A_{ij} ,\quad
 [\hat H^{(A)}_{\rm SHO},\hat B_{ij}]= -2\hbar\omega \hat B_{ij} .
\end{array}
\ee
Thus, the U(3) subalgebra is   spanned by the subset of operators 
$\{ \hat{\mathcal{Q}}_{ij},\hat L_{ij}\}$ that commute with 
$\hat H^{(A)}_{\rm SHO}$, and  the operators 
$\{ \hat A_{ij}\}$ and $\{ \hat B_{ij}\}$ are, respectively,   $\pm 2 \hbar\omega$ raising and lowering  operators for  representations of the sp$(3,\Rb)$) algebra.
The $L=0$ combination $\hat A_0 = \sqrt{\frac23}\sum_i \hat A_{ii}$
 is the creation operator for a $2\hbar\omega$ giant-monopole excitation and
the five $L=2$  combinations 
$\{ \hat A_{2\nu}; \nu = 0, \pm 1,\pm 2\}$  are creation operators for  
$2\hbar\omega$ giant-quadrupole excitations.
Together, these six raising operators transform as the $L=0$ and 2 components of an SU(3) (2\,0) tensor (see Exercise \ref{ex:7.20SU3tensor}).

\subsection{Irreducible representations (irreps) of {Sp(3,\,R)} in a U(3) basis}

The Sp(3,\,R) Lie alglebra is non-compact and, like the Lie algebra of the Bohr model, it has only  infinite-dimensional unitary irreps.  In the space of the nuclear shell model it has irreps with lowest weights, but none with highest weights.
Thus, irreps of Sp(3,\,R) are  constructed within the shell-model space by first defining their lowest-weight states which are also conveniently defined as U(3)  highest-weight states.

A U(3) highest-weight state is  labelled by three quantum numbers
$|N(\lambda \mu)\rangle$ and satisfies the equations
\bea
&&\hat C_{ii} |N(\lambda \mu)\rangle = N_i |N(\lambda \mu)\rangle , \quad i = 1,2,3,  \\
&&(\hat C_{11} + \hat C_{22} +\hat C_{33})|N(\lambda \mu)\rangle
=N |N(\lambda \mu)\rangle ,    \label{eq:U3h1}\\
&&(\hat C_{11} - \hat C_{22})|N(\lambda \mu)\rangle
=\lambda |N(\lambda \mu)\rangle, , \label{eq:U3h2}\\
&&( \hat C_{22}- \hat C_{33})|N(\lambda \mu)\rangle
=\mu |N(\lambda \mu)\rangle,  \label{eq:U3h3} \\
&&\hat C_{ij} |N(\lambda \mu)\rangle = 0, \quad \forall\, 1\leq i < j \leq 3, 
\label{eq:U3h4}
\eea
with
\be 
N=N_1+N_2+N_3,  \quad  \lambda = N_1-N_2, \quad \mu = N_2-N_3. \ee
An Sp(3,\,R) lowest-weight state $|N_0(\lambda_0 \mu_0)\rangle$ 
is a U(3) highest-weight state that is also annihilated by  the giant-resonance lowering operators, i.e.,
\be \hat B_{ij} |N_0(\lambda_0 \mu_0)\rangle = 0, \quad \forall\, 1\leq i\leq j\leq 3. \ee
A U(3) irrep with highest-weight state $|N(\lambda \mu)\rangle$ is denoted by
$\{N(\lambda \mu)\}$ and an Sp(3,\,R) irrep with lowest-weight state 
$|N_0(\lambda_0 \mu_0)\rangle$ is denoted by 
$\langle N_0(\lambda_0 \mu_0)\rangle$. A U(3) irrep whose highest-weight state is also an Sp(3,\,R) lowest-weight state is said to be a lowest-grade U(3) irrep for that Sp(3,\,R) irrep.
It follows that basis states for an Sp(3,\,R) irrep are obtained by augmenting a basis for a lowest-grade U(3) irrep by addition of an infinite set of giant-resonance excitations generated by the $\{ \hat A_{ij}\}$ raising operators.

There are four classes of U(3) irreps: $\{N(0\,0)\}$,
 $\{N(\lambda\, 0)\}$, $\{N(0\, \lambda)\}$ and 
 $\{N(\lambda\,\mu)\}$.
 The  $\{N(0\, 0)\}$ irreps are one-dimensional and contain a single $L=0$ state. 
 The $\{N(0\, \lambda)\}$ and $\{N(\lambda\, 0)\}$ irreps are  contragredient  to one another;  they have common spectra of angular momentum states, given by 
\be L= \lambda,\; \lambda-2,\; \lambda -4,\; \dots , 1\; {\rm or}\; 0 ,\ee
and differ only in that their quadrupole matrix elements are of opposite sign.  
A generic U(3) irrep  $\{N(\lambda\mu)\}$ contains a single SU(3) irrep 
$(\lambda\mu)$ and has a  spectrum of SO(3) angular-momentum states given by the  ${\rm SU}(3) \to {\rm SO}(3)$ branching  rule 
\bea  (\lambda\mu) &\mapsto&
      L= \lambda , \; \lambda-2  , \;  \dots  , \; 1\; {\rm or}\; 0\,,
        \qquad\, {\rm when} \; K= 0 , \nonumber\\
                               & \mapsto&
     L=K , \; K+1 , \;  \dots\, , \; K+\lambda,  
         \quad \quad \! {\rm when} \; K\not=0, \nonumber \\
& & \qquad {\rm for}\;\; K= \mu,\; \mu-2,\; \mu-4,\;\dots ,\; 1\; {\rm or}\; 0\,. 
\label{eq:6.SU3SO3brule}
\eea

\begin{figure}[ht]
\includegraphics[width=4.5in]{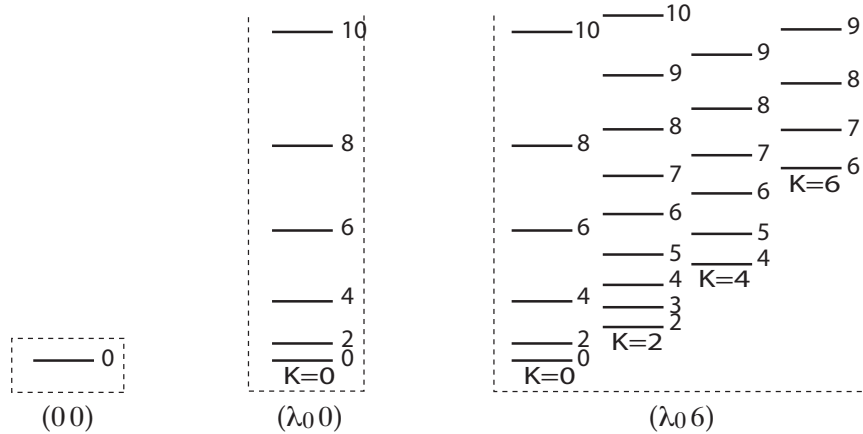}
\caption{Typical SU(3) model spectra for the lowest-energy states of three  irreps of even values of $\lambda_0$.  The (0 0) irrep contains a single state.
\label{fig:SU3irreps}}
\end{figure}

At this point it is useful to recall that  SU(3) irreps closely resemble truncated irreps of a rotor.  In fact,  for  values of $\lambda$ and/or $\mu$ large compared to the angular momenta of the  states of interest, 
the properties of the SU(3) irreps $(\lambda\, 0)$ and $(0\, \lambda)$ approach those of prolate and oblate rigid rotors, respectively, whereas the properties of a generic SU(3) irrep $(\lambda\mu)$ approach those of a triaxial rigid rotor.
Thus, an SU(3) model was introduced by Elliott \cite{Elliott58ab} as a first step towards a shell-model description of rotational states in light nuclei.  Typical spectra for SU(3) model irreps are shown in Fig,\ \ref{fig:SU3irreps}.

\begin{figure}[ht]
\includegraphics[width=5.7in]{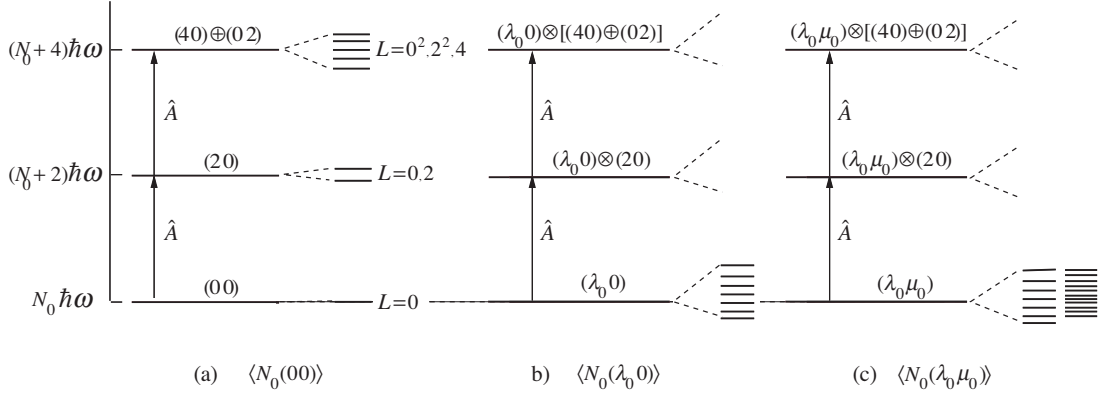}
\caption{Low basis states for Sp$(3,\Rb)$ irreps ordered by their spherical harmonic oscillator energies:  (a) for an $\langle N_0(0\,0)\rangle$ irrep; (b) for a  $\langle N_0(\lambda_0\,0)\rangle$ irrep;
(c) for a generic $\langle N_0(\lambda_0\,\mu_0)\rangle $ irrep.
\label{fig:Sp.bases}}
\end{figure}

The energy spectrum of basis states for an Sp(3,\,R) irrep, with respect to
 the spherical harmonic-oscillator Hamiltonian, Eq. (\ref{eq:Hsho}), is now obtained as follows.
The states of a lowest-grade U(3) irrep  
$\{\langle N_0(\lambda_0\,\mu_0)\rangle\}$ all have the common  
harmonic-oscillator energy $N_0 \hbar\omega$.  One-phonon giant-resonance states  appear at an energy of $(N_0+2)\hbar\omega$ and because the giant-resonance raising operators are components of an SU(3) (2\,0) tensor, they generate the states of a generally reducible SU(3) representation given by the SU(3) tensor product
$(\lambda_0\mu_0) \otimes (2\, 0)$.  The spectra of basis states are illustrated in Fig.\ \ref{fig:Sp.bases}.

\begin{figure}[ht]
\includegraphics[width=4in]{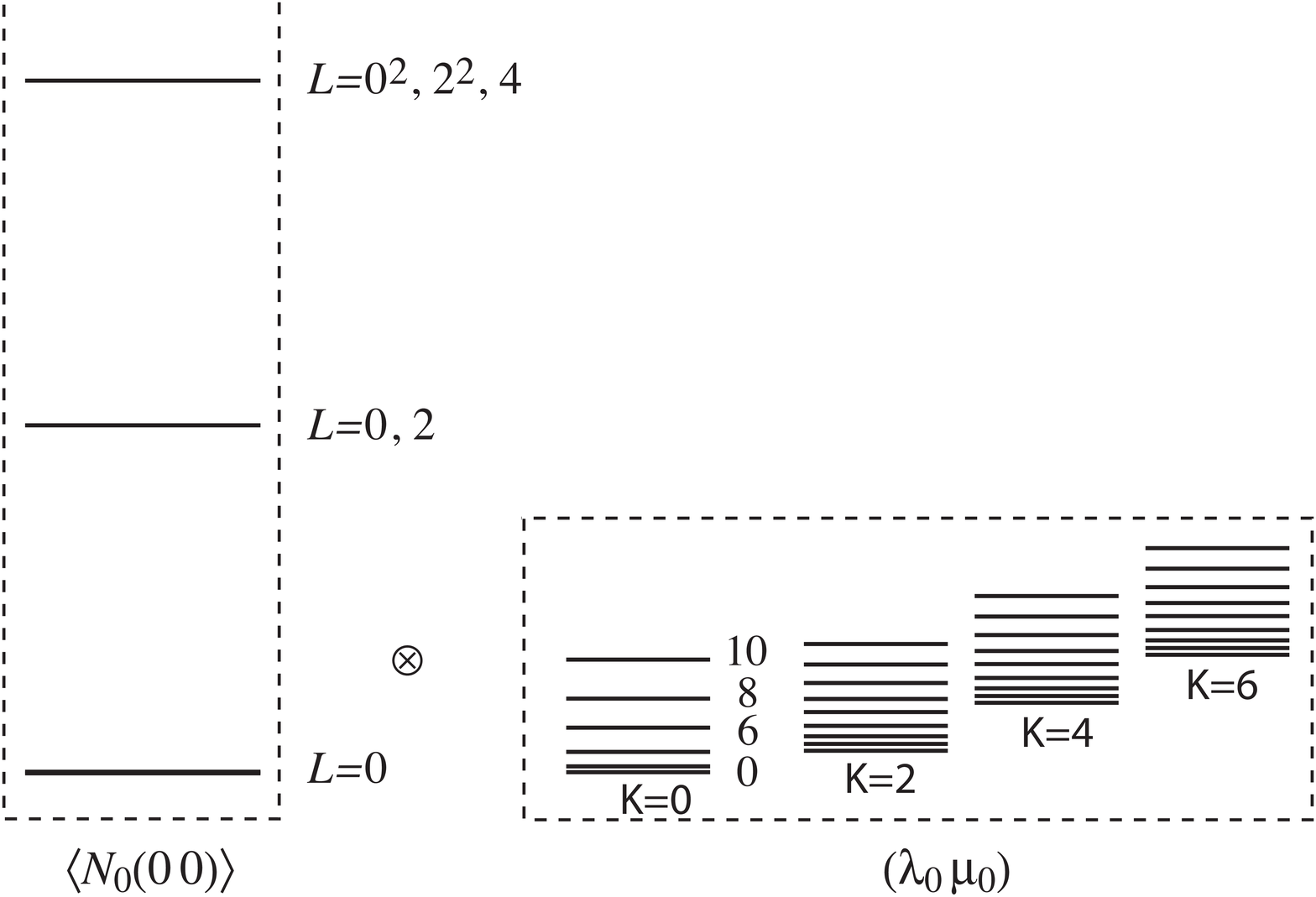}
\caption{Basis states for  the Sp(3,\,R) model are in 1-1 corresondence with a coupled product of the Bohr model (realised in terms of giant monopole-quadrupole resonance states with irrotational flows) and an SU(3) model.
\label{fig:BMSU3}}
\end{figure}

It is of interest to note that the spectrum of states for an 
$\langle N_0(0\,0)\rangle$ irrep is in 1-1 correspondence with that of a 6-dimensional harmonic oscillator.
In fact, they are a basis for an irrep of the microscopic Bohr model, spanned by the  $\{ \hat Q_{ij} , \hat P_{ij}\}$ operators of Eq.\ (\ref{eq:P&Q}),
which comprises  giant monopole and quadrupole harmonic vibrational states with irrotational-flow mass parameters.
More generally, the states of an Sp(3,\,R)  irrep are  in 1-1 correspondence with the states of such a Bohr model  coupled to a non-trivial SU(3) irrep.
This is illustrated pictorially in Fig.\ \ref{fig:BMSU3}.
Moreover, it has been shown \cite{RosensteelR81,LeBlancCR84,RoweVC89}
that, in the limit of large quantum numbers, the Sp(3,\,R) model  contracts to the coupled product of a Bohr vibrational model and a rotor   model.  
This is the so-called coupled rotor-vibrator limit of the symplectic model.

A simple prediction that follows, without a calculation, from the observation that the underlying microscopic description of monopole/quadrupole rotational and vibrational dynamics is provided by the symplectic model, is that this model contains no low-energy $\beta$ or $\gamma$ vibrational bands 
\cite{CarvalhoLeBRMcG86}.  It does exhibit excited $K=2$ bands but these are more of a triaxial rotor than of a vibrational nature.  Excited $K=0$ bands (for even-even) nuclei do occur in the model but are associated with an excited Sp(3,\,R) irrep.  With mixing of Sp(3,\,R) irreps, such an excited band can exhibit properties associated with a $\beta$ band.  However, the occurrence of such bands
is much more appropriately described in terms of band mixing and as a departure from the underlying symmetry rather than as a part of it.  The study of such $K=0$ bands and the E2 transitions between them is therefore expected to provide valuable information on the mixing of Sp(3,\,R) irreps in even-even nuclei
(see, e.g., \cite{HeydeW11}).

\subsection{A model Sp(3,\,R) calculation}

\begin{figure}[ht]
\includegraphics[width=5in]{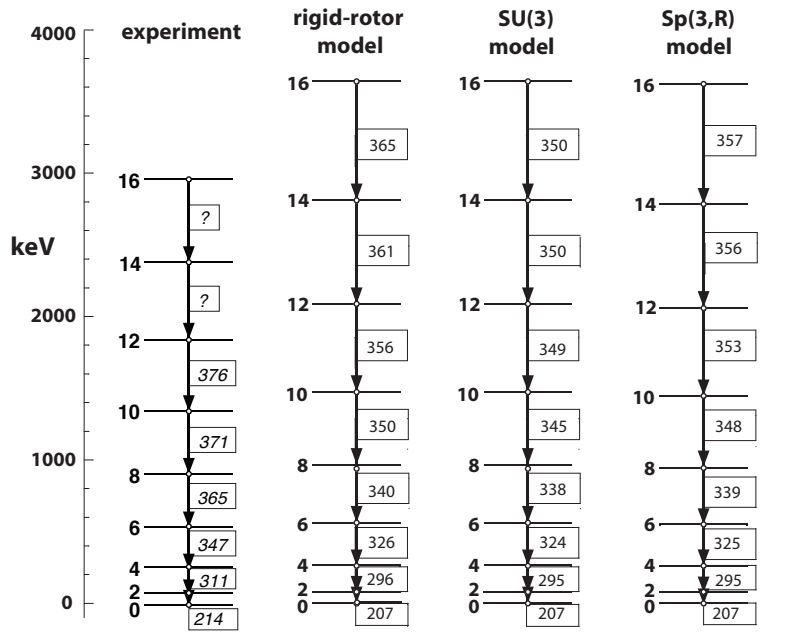}
\caption{Low-energy states of $^{166}$Er calculated in an Sp(3,\,R) irrep
$\langle 826.5(78,0)\rangle$ for the Hamiltonian (\ref{eq:HSp3R}).
 For comparison the spectra calculated in the rotor and SU(3) models are also shown. The boxed numbers are B(E2)'s in W.u.
\label{fig.Sp3R166Er}}
\end{figure}

Figure \ref{fig.Sp3R166Er} shows the energy  levels and reduced E2 transition rates for a symplectic model calculation \cite{BahriR00} for an Sp(3,\,R) irrep
$\langle 826.5(78,0)\rangle$ with the two-parameter Hamiltonian
\be \hat H = \hat H^{(166)}_{\rm SHO} + \chi\left( \hat Q\cdot \hat Q+ \frac{\varepsilon}{\hat Q\cdot \hat Q}\right).
\label{eq:HSp3R}
\ee
The  irrep $\langle 826.5(78,\,0)\rangle$  was chosen  to be appropriate for 
$^{166}$Er on the basis of  experimental data \cite{JarrioWR91} (as discussed below).
The two parameters $\varepsilon$ and $\chi$ of this Hamiltonian are naturally defined as follows:
$\varepsilon$ can be adjusted so that the potential energy component of the Hamiltonian (\ref{eq:HSp3R}) has a minimum at the  deformation  determined  from experimental E2 transition rates;  the strength parameter $\chi$ is then determined by the self-consistency condition that the density distribution for the wave function has the same quadrupole shape as that of the potential.
For the results shown, the simpler procedure was followed of adjusting 
$\varepsilon$ as mentioned above to accord with the observed deformation, and adjusting $\chi$ to give the observed energy of the $2_1$ state.
The close agreement of the results, especially for states of lower angular momentum is very satisfying because it means that the symplectic model is describing the physics of a rotational band of states with realistic moments of inertia that emerge naturally without any adjustment of the many-nucleon kinetic energy, e.g., by introducing an effective nucleon mass.

The fact that the symplectic model agrees  closely with a two-parameter rotor model is also satisfying because, while the rotational character of the states described is presumed in the rotor model, no such  assumption is  made in the symplectic model which has many  more degrees of freedom.  In fact, it is apparent that the experimental energy levels exhibit a degree of centrifugal stretching that is not admitted in the rigid-rotor model.   Thus,  the rotor model interpretation of the symplectic model results suggests that a better agreement with the higher energy levels could have been obtained in the symplectic model with a potential energy having a less rigidly-defined minimum.
The agreement between the two-parameter SU(3) and rigid-rotor models is not surprising, in view of the contraction, noted above, of the SU(3) model to the rotor model for large-dimensional representations.  
However, the relationship between the symplectic model and these two models
is important for understanding the microscopic structure of rotational states and for understanding how the seeds of nuclear rotations can be found within the framework of a shell-model coupling scheme.

\section{The shell-model theory of deformed nuclei}

Because the many-nucleon Hilbert space is infinite dimensional, many-nucleon calculations must necessarily be restricted to finite-dimensional subspaces.  Thus, to obtain meaningful results, it is necessary to choose the finite-dimensional subspaces such that they have significant relationships with the physical states they purport to describe.

The standard shell model is based on the presumption that the dominant forces on a nucleon in the nucleus are generated by a spherically symmetric mean-field with spin-orbit interactions.  Thus, a hierarchy of standard shell model states is  defined by  a partially-ordered basis of  independent-particle model states.
In this model, many-nucleon subspaces are naturally defined by specification of occupied and unoccupied single-particle states and a complementary valence space of partially-occupied states.
Such subspaces can be ordered by their summed independent-particle-model energies.
Which subspace one chooses for a specific purpose obviously depends on one's objectives and computational resources.
However, a common practice is to choose a subspace that is as large as can be handled and, apart from symmetry constraints (such as parity), is one for which all independent-particle states of energy below those of the valence space are occupied.

In practice, the mean field of the shell model is  commonly replaced by a spherical harmonic oscillator potential.  This makes it possible to capitalise on the rich variety of symmetries associated with the harmonic oscillator, which greatly facilitate the execution of complex calculations.  It must be  remembered though that the purpose of the independent-particle potential is to define basis states for a many-nucleon calculation and the better the basis the fewer basis states are needed to obtain results to a given level of accuracy.  Except for the radial tails of the wave functions, which are  important in reaction theory but less important for the calculation of bound states, a spherical harmonic oscillator does a reasonably good job of approximating the mean field  for spherical nuclei and the inadequacies of the bases that it provides are more than compensated by the many benefits resulting from the harmonic oscillator symmetries.

A concern addressed in these lectures is that, while the standard spherical harmonic oscillator shell model is very successful for the description of spherical nuclei, it fails badly in the description of strongly-deformed nuclei.
This is revealed  by considering which Sp(3,\,R) irreps are appropriate for the description of rotational bands in nuclei and finding that they lie in shell-model subspaces that are often far above those that would normally be included in the valence spaces of the spherical harmonic-oscillator shell model. It is insightful to consider this concern from a mean-field perspective.

\subsection{Geometrical and mean-field perspectives}

Loosely speaking, mean-field theory can be regarded as quantum mechanics constrained to an orbit of a dynamical group that is isomorphic to a classical phase space.
Thus, mean-field theory provides an interface between classical and quantum mechanics and enables one to obtain valuable classical insights into the quantum world.  
For example, in quantum mechincs, the ground state of a Hamiltonian is contained in an irrep of a dynamical group for that Hamiltonian.  In mean-field theory, a classical ground state  is an equilibrium state given by the minimal-energy state on the classical phase space.
This minimum energy is then an upper bound on the quantum-mechanical ground state energy.
Thus, it is meaningful to order Sp(3,\,R) irreps by their minimal energy mean-field states.


Let $|\sigma\rangle = |N_0(\lambda_0\mu_0)\rangle$ denote a lowest-weight state for an Sp(3,\,R) irrep and let $\hat T(g)$  denote a unitary transformation of the states of this irrep by an element $g\in {\rm Sp(3,\,R)}$.  The group orbit is then  the manifold of coherent states
\be {\cal M} = \{|\sigma(g)\rangle = \hat T(g) |\sigma\rangle; \; g\in {\rm Sp(3,\,R)}\} ,\label{eq:calMSp3R}\ee
which is isomorphic to the classical phase space for the irrep.
We now claim that the real submanifold of zero-momentum states of this phase space is  the set
\be {\cal R} = \{|\sigma(g)\rangle = \hat T(g) |\sigma\rangle; \; g\in {\rm GL_+(3,R)}\} , \label{eq:realSp3Rcohstates}\ee
where ${\rm GL_+(3,R)}$ is the connected subgroup of real $3\times 3$ matrices with positive determinant.
To see this, first observe that the  phase space ${\cal M}$ has dimension $\dim({\cal M})= 18$; this follows because the Sp(3,\,R) Lie algebra is of dimension 21 but, as shown by Eqs.\
(\ref{eq:U3h1}) -- (\ref{eq:U3h3}), three elements of this Lie algebra, namely
$\hat C_{11}, \hat C_{22}$ and $\hat C_{33}$, leave the state $|\sigma\rangle$ invariant.  Next observe that the nine linearly-independent momentum operators
$\hat P_{ij}= \hat P_{ji}$ and $\hat L_{ij}= - \hat L_{ji}$ are the infinitesimal generators of the real subspace ${\cal R} \subset {\cal M}$ and that they span a Lie algebra isomorphic to that of ${\rm GL_+(3,R)}$. 

We now make use of the fact that a general linear matrix $g\in {\rm GL_+(3,R)}$ can be expressed as a matrix product
\be g = \Omega {\bf d} \Omega' ,\ee
where $\Omega$ and $\Omega'$ are SO(3) matrices and 
${\bf d}= {\rm diag}(d_1,d_2,d_3)$ is a real diagonal matrix with positive entries.
Thus, in the unitary representation $\hat T$, we can express the mean-field states
of ${\cal R}$ in the form
\be |\sigma(\Omega d \Omega')\rangle =
\hat R(\Omega) \hat U({\bf d}) \hat R(\Omega') |\sigma\rangle ,\ee
where $\hat R(\Omega)$ is a rotation/inversion operator and $\hat U({\bf d})$ is a unitary representation of the scale transformation
$\{x_{ni} \to d_i x_{ni},\, i=1,2,3\}$.

\begin{figure}[ht]
\centerline{\includegraphics[width=5in]{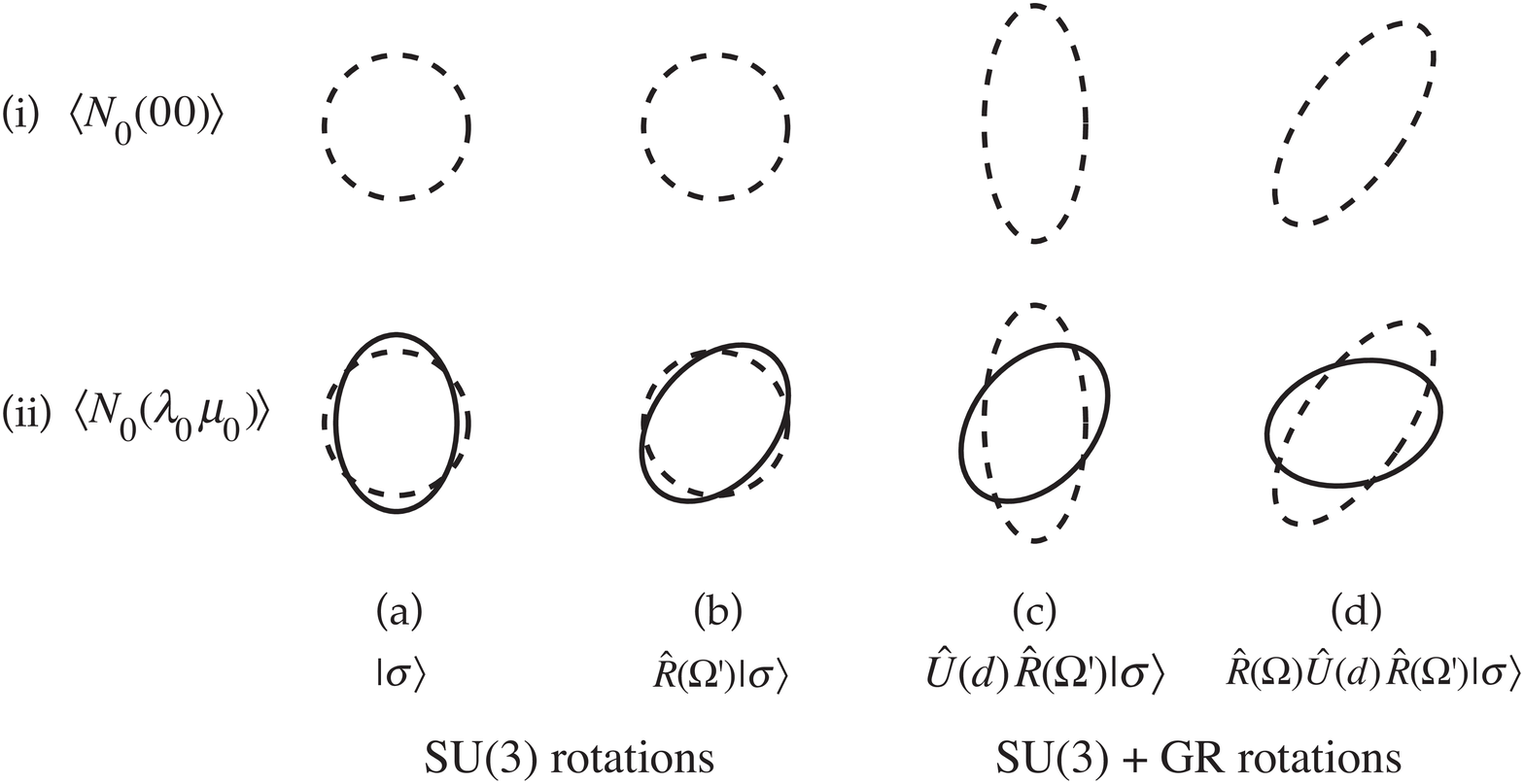}}
\caption{A geometrical perspective of the nine dynamical degrees of freedom of the symplectic model which reduce to  six   for a spherical irrep 
$\langle  N_0(0\, 0)\rangle$ wherein the three SU(3) rotational degrees of freedom are suppressed; see the text for an explanation of this figure.
\label{fig:Sp3Rgeom}}
\end{figure}

The sequence of transformations 
$\hat R(\Omega) \hat U({\bf d}) \hat R(\Omega')$ of the Sp(3,\,R) lowest-weight state $|\sigma\rangle$ is represented pictorially in 
Fig.\ \ref{fig:Sp3Rgeom} for two Sp(3,\,R) irreps, $\langle N_0 (0\, 0)\rangle$ and a generic irrep $\langle N_0(\lambda_0\mu_0)\rangle$.
For the purpose of this representation, it is convenient to make use of the 1-1 correspondence between the basis states of the symplectic model and those of a coupled Bohr (giant-resonance) model and an SU(3) model as depicted in Fig.\ \ref{fig:BMSU3}.
The  states of the manifold ${\cal R}$ can then be represented by a pair of ellipsoids: one of fixed shape but arbitrary orientation, representing the three rotational degrees of freedom of the SU(3) rotor; and one with a full range of ellipsoidal shapes and orientations, representing the six degrees of freedom of the Bohr model. The states of the SU(3) rotor  are drawn in the figure with solid lines while  those for the giant-resonance are drawn with dashed lines.
Recall, however, that for the Sp(3,\,R) irrep $\langle N_0 (0\,0)\rangle$ the lowest-grade SU(3) irrep is the trivial identity representation.  For this irrep only the giant-resonance degrees of freedom survive and the lowest-weight state 
$\langle N_0 (0\,0)\rangle$ is the giant-resonance vacuum state which has a spherical density distribution as shown in row (i) column (a) of the figure.
For a generic Sp(3,\,R) irrep the lowest-weight state $|\sigma\rangle$ is again a giant-resonance vacuum state.  But it is also the highest-weight state for a non-trivial SU(3) irrep with a non-spherical density distribution, as shown in row (ii) column (a).


Because the action of the rotation operator $\hat R(\Omega')$ on the lowest-weight state $|\sigma\rangle$ does not excite any giant-resonance states, its effect is simply to reorient the SU(3) ellipsoidal density distribution.  
Conversely, the  transformation $\hat U({\bf d})$, shown in (c), is a diagonal scale transformation which  excites giant resonance degrees of freedom and  changes the spherical shape of the giant-resonance vacuum state to a more general ellipsoidal shape, with its principal axes aligned along the space-fixed axes, but  leaves the SU(3) degrees of freedom alone.
Finally the last rotation $\hat R(\Omega)$, shown in (d) rotates both the SU(3) and giant-resonance ellipsoids.
The resulting shape of the nucleus, as represented by 
Fig.\ \ref{fig:Sp3Rgeom}(d), is now  a composition of the two ellipsoids shown in which that corresponding to the SU(3) lowest grade ellipsoid is scaled in a manner defined by the GR ellipsoid.

\subsection{The minimal-energy mean-field state for an Sp(3,\,R) irrep}

To derive the minimal-energy mean-field state for an Sp(3,\,R) irrep, one would ideally make a variational calculation, analogous to a
 Hartree-Fock calculation, on the phase space of Sp(3,\,R) coherent states defined by Eq.\ (\ref{eq:calMSp3R}).
However, it is possible to  assess what the result would be on the basis of a simple self-consistency argument.
First observe that, for a time-reversal-invariant Hamiltonian, the minimal-energy state should be one with vanishing expectation values of its momentum operators.
Thus, it should be the state in the  set 
$\{ |\sigma(g)\rangle ,\, g\in {\rm GL_+(3,R)}\}$
for which 
\be E( \Omega {\bf d} \Omega')=
\langle \sigma (\Omega {\bf d} \Omega')| \hat H |\sigma ( \Omega{\bf d} \Omega') \rangle  \label{eq:E(OmdOm'}\ee 
is a minimum for the given Hamiltonian.

The next observation is that, for a rotationally-invariant Hamiltonian, the energy is independent of $\Omega$.
We  also suppose that, because the giant-resonance
contribution to the deformation of a minimal-energy state arises from core polarisation of the nucleus by the quadrupole field of the SU(3) states, the minimal-energy state should be one for which the polarisation is maximally aligned with the SU(3) deformation.  Inspection of Fig.\ \ref{fig:Sp3Rgeom} shows that such an alignment is achieved when the rotational angle $\Omega'$ is zero.
It then remains to determine the scale deformation for which $E({\bf d})$ is a minimum. 

We start from the following observations:
\begin{enumerate}
\item[(i)]   By construction, the lowest-weight state 
$|\sigma\rangle = |N_0 (\lambda_0\mu_0)\rangle$ is an eigenstate of an
isotropic harmonic-oscillator Hamiltonian
\be \hat H_0 = \sum_n\left[\frac{\hat{\bf p}_n^2}{2M} + 
\frac12 M \omega^2 \left( x_{n1}^2 + x_{n2}^2+ x_{n3}^2\right) \right] 
\label{eq:SHO} \ee
with eigenvalue given by
\be \hat H_0 |\sigma\rangle = N_0 \hbar\omega |\sigma\rangle . \ee
\item[(ii)]  Every state $|\sigma(g)\rangle\in {\cal R}$ is a lowest-weight state for an Sp$(3,\Rb)$ irrep with respect to some harmonic-oscillator Hamiltonian.
\item[(iii)]  The lowest-weight state $|\sigma({\bf d})\rangle$ is an eigenstate of a triaxial harmonic oscillator
\be \hat H_{\bf d} = \sum_n\left[\frac{\hat{\bf p}_n^2}{2M} + 
\frac12 M  \left(\omega_1^2 x_{n1}^2 + \omega_2^2 x_{n2}^2 +
\omega_3^2 x_{n3}^2\right) \right] . \label{eq:DHO} \ee
\end{enumerate}

We now determine the frequencies $\{ \omega_i\}$ of this triaxial oscillator for the minimal-energy state by use of the self-consistency property of mean-field theory  which is that the minimum-energy  state should be one for which (to the extent allowed by the mean-field constraints) the equidensity surfaces of the state have the same shape as the corresponding equipotential surfaces of the independent-particle Hamiltonian of which it is an eigenstate.

It is important to understand that, although the lowest weight $|\sigma\rangle$ is  observed to be an eigenstate of an independent-particle Hamiltonian, there is no implication that $|\sigma\rangle$ is an independent-particle state.  However, in many important cases, e.g., for an Sp$(3,\Rb)$ irrep built on a leading SU(3) irrep, it may be.

An ellipsoidal surface associated with the density distribution of the lowest-weight state  $|\sigma\rangle$ is defined by the expectation values
\be
 \langle x_i^2\rangle_\sigma = \langle \sigma | \sum_n x_{ni}^2 |\sigma\rangle 
 = \frac{\hbar N_i}{M\omega}\, , \quad i=1,2,3, \ee
and, unlike the isotropic  harmonic-oscillator Hamiltonian (\ref{eq:SHO}) for which this state is an eigenstate, it is not spherically symmetric.
On the other hand, a state  $|\sigma(d)\rangle$ which is an eigenstate of an anisotropic harmonic oscillator with potential
\be V(x) = \frac12 M \sum_n \left(\omega_1^2 x_{n1}^2 + \omega_2^2 x_{n2}^2 + \omega_3^2 x_{n3}^2\right)  , \label{eq:V(x)}\ee
has the mean values 
\be
 \langle x_i^2\rangle_{\sigma({\bf d})} 
 = \langle \sigma({\bf d}) | \sum_n x_{ni}^2 |\sigma({\bf d})\rangle 
 = \frac{\hbar N_i}{M\omega_i}, \ee
An ellipsoidal equi-density surface for this state, defined by the equation
\be \frac{x_1^2}{\langle x_1^2\rangle_{\sigma({\bf d})}}
+ \frac{x_2^2}{\langle x_2^2\rangle_{\sigma({\bf d})}}
+ \frac{x_3^2}{\langle x_3^2\rangle_{\sigma({\bf d})}} = {\rm const.}
\ee
is then given by 
\be \frac{\omega_1x_1^2}{N_1} + \frac{\omega_2x_2^2}{N_2} 
 +\frac{\omega_3x_3^2}{N_3}= 1   \label{eq:equidensity}\ee
whereas an equi-potential surface for the potential (\ref{eq:V(x)}) is given by
\be \omega_1^2 x_1^2 + \omega_2^2 x_2^2 +\omega_3^2 x_3^2 
= 1. \label{eq:equipot}\ee
Thus, for these surfaces  to have the same shape, it is required that 
\be N_1\omega_1 = N_2\omega_2 =N_3\omega_3 =k , \label{eq:SCcondition} \ee
for some value of $k$.

The volume of the ellipsoid (\ref{eq:equidensity}) is
\be {\rm vol.} = \frac{4\pi}{3}
\left( \frac{N_1N_2N_3}{\omega_1\omega_2\omega_3} \right)^{\! \frac{1}{2}} .
\ee
Thus, in order that this volume should be the same  as that for 
$\omega_1=\omega_2=\omega_3=\omega$, we set
\be \omega_1\omega_3\omega_3 =\omega^3\ee
and obtain
\be k^3 = N_1N_2N_3 \omega^3. \ee
This means that the scale parameters for the minimal-energy state should be such that
\be  \frac{\langle x_i^2\rangle_{\sigma(d)}}{\langle x_i^2\rangle_\sigma} =  
\frac{\omega}{\omega_i} = \frac{N_i\omega}{k}= \frac{N_i}{(N_1N_2N_3)^{1/3}} .
\label{eq:shapecondns} \ee

The above self-consistency method extends to all orders the methods  used in Refs.\ \cite{Rowe67,LeBlancCVR86,RoweTW06}
 to estimate the strength 
$\chi = \hbar\omega/N_0$ to leading order in 
$\lambda/N_0$ and $\mu/N_0$,  of the  coupling constant for the effective quadrupole-quadrupole interaction in a model Hamiltonian  of the form
$\hat H = \hat H_{\rm sp} - \frac12 \chi Q\cdot Q$.

Several important consequences  follow from this shape-consistency result.
\begin{enumerate}
\item[(i)]
\emph{Coupling to giant resonance states (essentially) doubles the values of 
SU(3) quadrupole moments.}\smallskip

This follows because quadrupole moments of the state $|\sigma\rangle$ are given in the SU(3) model in a lowest-grade irrep, in terms of the oscillator unit $a= \sqrt{M\omega/\hbar}$, by
\bea
&& a^2\langle (2x_1^2 - x_2^2-x_3^2) \rangle_\sigma = 2N_1-N_2-N_3 
 =2\lambda_0+\mu_0 , \\
&& {\textstyle a^2 \sqrt{\frac23}}\langle  (x_2^2-x_3^2) \rangle_\sigma = N_2-N_3 = \mu_0,
\eea
Therefore, for the shape-consistent state $|\sigma({\bf d})\rangle$, they are given by
\bea
&& a^2\langle 2x_1^2 - x_2^2-x_3^2\rangle_{\sigma({\bf d})} 
= \frac{2N^2_1-N^2_2-N^2_3}{(N_1N_2N_3)^{1/3}} ,  \\
&& \textstyle  a^2 \sqrt{\frac23} \langle x_2^2-x_3^2\rangle_{\sigma({\bf d})} 
=\displaystyle \frac{N^2_2-N^2_3}{(N_1N_2N_3)^{1/3}}  .
\eea
An expansion in powers of $\lambda_0/N_0$ and $\mu_0/N_0$ then gives 
\bea
&& a^2\langle 2x_1^2 - x_2^2-x_3^2\rangle_{\sigma({\bf d})} 
=   2(2 \lambda_0+ \mu_0) + (2\lambda_0^2+2\lambda_0\mu_0 - \mu_0^2)/N_0 + \dots ,  \\
&& {\textstyle a^2 \sqrt{\frac23}} \langle x_2^2-x_3^2\rangle_{\sigma({\bf d})} 
= 2\mu_0 - (2\lambda_0\mu_0 - \mu_0^2)/N_0 + \dots ,
\eea
which, to leading order in  $\lambda_0/N_0$ and $\mu_0/N_0$, imply that the effective charge of quadrupole moments and E2 transitions in the SU(3)  model is predicted to be
\be e_{\rm eff} = \big(2 +{\rm higher\; order\; terms.}\big) \times
{\rm the\; bare\; nucleon\; charge}, \ee
where the bare  nucleon charge for an SU(3) irrep would normally be 
$Z/A$ times the proton charge.

\item[(ii)]
\emph{Coupling to the giant resonance excitations lowers the energies of deformed states.}\smallskip

To estimate the magnitude of this lowering, one should compare the expectation values of
$\langle \sigma | \hat H |\sigma\rangle$ and
$\langle \sigma({\bf d}) | \hat H |\sigma({\bf d})\rangle$ for a realistic nuclear Hamiltonian.
However, in the spirit of the shell model in which a first energy-ordering of states is given by their harmonic oscillator energies, we consider the energy of the state $|\sigma(d)\rangle$ for the shape-consistent harmonic oscillator Hamiltonian
$\hat H_d$ of Eq.\ (\ref{eq:DHO}).
This gives
\be E_{N_1,N_2,N_3}=\langle \sigma({\bf d}) | \hat H_{\bf d} |\sigma({\bf d})\rangle
= 3\hbar k = 3(N_1N_2N_3)^{1/3} \hbar \omega , \ee
which is less than the energy of the lowest-weight state for the same irrep relative to the spherical harmonic oscillator Hamiltonian 
\be E_{N_0} =N_0  \hbar \omega= (N_1+N_2+N_3) \hbar \omega .\ee
A consequence of this result is:

\item[(iii)]
\emph{The lowest-energy shell-model states of well-deformed nuclei are unlikely to lie in the lowest-energy spherical-harmonic oscillator shells.}\smallskip

This claim is based on the observation that an Sp(3,\,R) irrep is defined uniquely by the quantum numbers $N_1$, $N_2$, and $N_3$ for which
\be N_0 = N_1+ N_2 +N_3, \quad \lambda_0 = N_1-N_2, 
\quad \mu_0 = N_2=N_3 .\ee
Moreover, all the mean-field states $\{ |\sigma(g)\rangle ,\; g\in {\rm Sp(3,\,R)}\}$ for a given Sp(3,\,R) irrep have common values for these quantum numbers and 
are of lowest-weight for the Sp(3,\,R) irrep with respect to some (generally triaxial) harmonic oscillator Hamiltonian.
We have argued that, with respect to any reasonable nuclear Hamiltonian, the lowest energy mean-field state should be one for which there is shape consistency between the density of the minimal-energy state and the harmonic oscillator Hamiltonian for which the minimal-energy state is an eigenstate.
A property of standard Hartree-Fock theory, on manifolds of Slater determinants, is that the minimal-energy state with respect to a given nuclear Hamiltonian is also of minimal energy with respect to the locally defined Hartree-Fock independent-particle Hamiltonian.  If we assume this property to hold for the above described Sp(3,\,R) mean-field theory, for which the analogue of the Hartree-Fock independent-particle Hamiltonian is the anisotropic harmonic oscillator Hamiltonian for which a given lowest-weight Sp(3,\,R) state is an eigenstate, the  minimal energy state with respect to the given nuclear Hamiltonian should also be a minimal energy state with respect to the locally-defined harmonic-oscillator Hamiltonian.\medskip

As noted above, the  lowest-weight state $|\sigma\rangle$ of an Sp(3,\,R) irrep 
$\langle N_0 (\lambda_0\mu_0)\rangle$ with respect to an isotropic harmonic oscillator Hamiltonian is an eigenstate of this Hamiltonian with energy 
$N_0 \hbar\omega$.  However, the shape-consistent lowest-weight state 
$|\sigma(d)\rangle$ of the same volume is an eigenstate of an anisotropic harmonic oscillator of energy eigenvalue 
$E_{N_1,N_2,N_3}=3(N_1N_2N_3)^{1/3} \hbar \omega$,
which is less than $N_0 \hbar\omega$ unless $N_1$, $N_2$, and $N_3$ are all equal.
Thus, it is appropriate to compare the minimal energies $E_{N_1,N_2,N_3}$ for the possible Sp(3,\,R) irreps that are available to a given nucleus.

\end{enumerate}

\subsection{Identifying the lowest-energy Sp(3,\,R) irreps}

The above shape-consistent mean-field results suggest that an energy ordering of the Sp(3,\,R) irreps available to a given nucleus is given by the values of
$E_{N_1,N_2,N_3}$.  For the given nucleus, we can list the possible values of $N_1$, $N_2$, and $N_3$ together with the corresponding values of
$E_{N_1,N_2,N_3}$.  The first few irreps in order of increasing
$E_{N_1,N_2,N_3}$ are shown for  $^{12}$C, $^{16}$O  and $^{168}$Er in 
Table \ref{eq:tab.1}.\\

\begin{table}[h]
\caption{Minimal mean-field energies $E_{N_1N_2N_3}$ in units of 
$\hbar\omega$ for the lowest-energy irreps of $^{12}$C, $^{16}$O, and 
$^{168}$Er as defined in the text. \vspace{0.2cm} \label{eq:tab.1}}

  \begin{minipage}[t]{1.5in}
\centerline{${}^{12}${\rm C} }
$\begin{array}{|c|c|c|c|c|}\hline
{N_0} &  \lambda_0 &  \mu_0 & E_{N_1N_2N_3}   \\ \hline
{ 26}  &  { 0}   &  { 4} &  { 25.3} \\ 
30  &    12   &  0 &  26.0 \\ 
28  &    6   &  2 &  26.3 \\ 
32  &    10   &  2 &  28.6 \\ 
\hline
\end{array}$
\end{minipage}   \qquad 
\begin{minipage}[t]{1.5in}
\centerline{${}^{16}${\rm O} }
$\begin{array}{|c|c|c|c|c|}\hline
{N_0} &  \lambda &  \mu & E_{N_1N_2N_3}  \\ \hline
{ 36}  &    { 0}   &  { 0} &  { 36} \\ 
{ 40}  &    { 8}    & { 4}  &  { 37.3}  \\ 
38  &    4   &  2 &  37.3 \\ 
40  &    7   &  3 &   38.1 \\ 
48  &    24   &  0 &  38.1 \\ 
44  &   16  &  2 &   38.3 \\ 
\hline
\end{array}$
\end{minipage} \qquad 
\begin{minipage}[t]{1.7in}
\centerline{${}^{168}${\rm Er} }
$\begin{array}{|c|c|c|c|c|}\hline
{N_0} &  \lambda &  \mu &  E_{N_1N_2N_3}   \\ \hline
814  &   30   &   8  &  812.6 \\ 
{ 826}  &   { 96}   &  { 20} &  { 812.9} \\
822  &   70   &  28 &  813.0 \\ 
818  &   52   &  20 & 813.1 \\
816  &   42   &  12 &  813.1 \\ 
830  & 114   &  16 &813.2 \\ 
\hline
\end{array}$
\end{minipage}
\end{table}

\begin{figure}[ht]
\centerline{\includegraphics[width=4in]{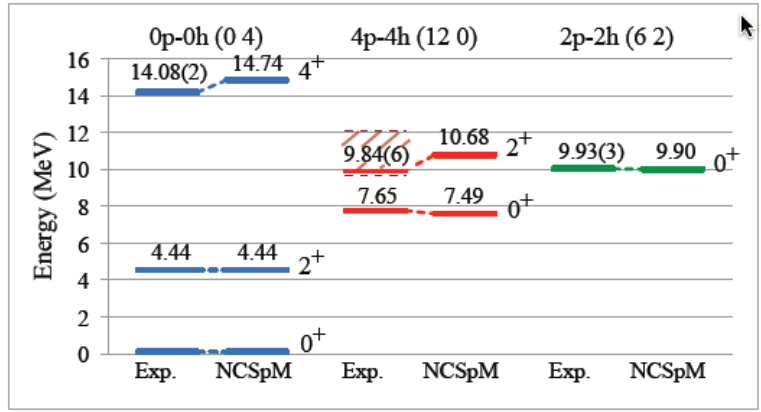}}
\caption{The low-energy spectrum of $^{12}$C energies calculated for the first three Sp(3,\,R) irreps shown for $^{12}$C in Table \ref{eq:tab.1} by Dreyfuss \emph{et al.} \cite{DreyfussLDDB12}. (The figure is from their preprint with permission.)
\label{fig:Hoyle}}
\end{figure}

\begin{figure}[hbtp]
\centerline{\includegraphics[width=5in]{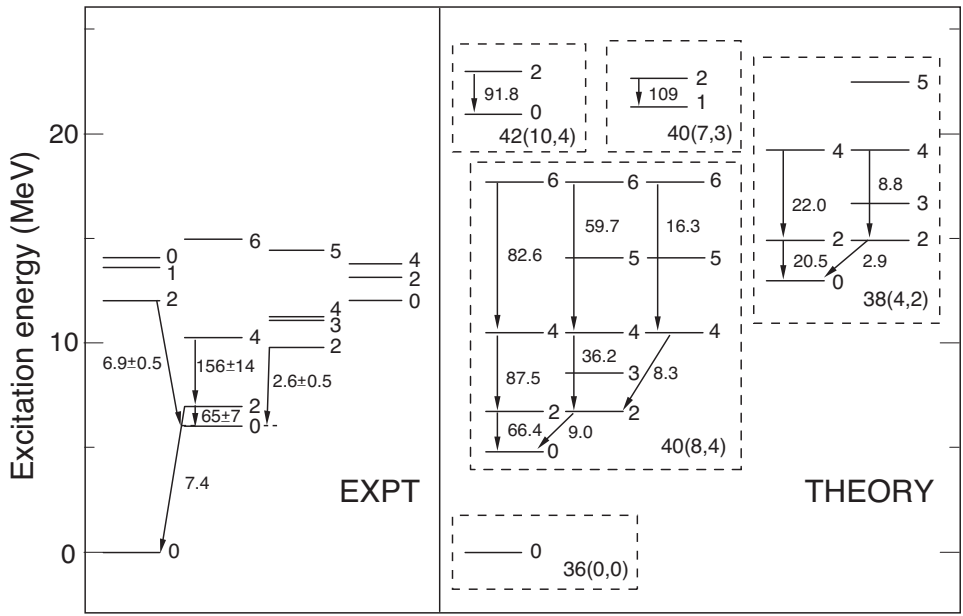}}
\caption{The low-energy spectrum of $^{16}$O energies calculated for a few Sp(3,\,R) irreps. (The figure is from Ref.\ \cite{RoweTW06}.)
\label{fig:16O}}
\end{figure}

The first remarkable result is that for the light nucleus $^{12}$C  the lowest  three 
Sp(3,\,R) irreps are consistent with the states observed in this nucleus as described in an Sp(3,\,R) model framework by Dreyfuss \emph{et al.}
\cite{DreyfussLBDD11,DreyfussLDDB12} whose results are shown in Fig.\ \ref{fig:Hoyle}. In particular, the highly deformed second excited state, known as the Hoyle state, which belongs to the $(\lambda_0, \mu_0)= (12\; 0)$ irrep, falls well below its unperturbed energy of $4\hbar\omega$  in the spherical harmonic oscillator shell model.
A similar, result is found for $^{16}$O, for which the lowest three irreps are consistent with those observed and described \cite{RoweTW06} in an Sp(3,\,R) model framework  in Fig.\ \ref{fig:16O}. For $^{16}$O, the first-excited state belongs to an irrep whose lowest-weight state would again be $4\hbar\omega$ above its unperturbed spherical harmonic oscillator and would lie outside of any normal shell-model valence space.

For  a range of Erbium and Tungsten isotopes, Jarrio \emph{et al.} \cite{JarrioWR91} estimated the most appropriate Sp(3,\,R) irreps needed to describe the ground-state rotational bands of these nuclei directly from experiments.  Their method was simply to fit the E2 transition-rate data with an Sp(3,\,R) irrep in its rotor-model limit.
They determined, for example, that the mean values of  $\lambda_0$ and 
$\mu_0$, for the low-energy states of $^{168}$Er  are  approximately 
$\langle\lambda_0\rangle \approx 80$ and $\langle \mu_0\rangle \approx 12$.  They also determined the mean values of these quantum numbers for Nilsson model states at the experimentally determined deformation and obtained the values $\langle\lambda_0\rangle \approx 95$ and 
$\langle \mu_0\rangle \approx 12$.  Given that there is undoubtedly a mixing of Sp(3,\,R) irreps due to symmetry breaking interactions, their results are reasonably consistent with the dominant Sp(3,\,R) irrep contributing to the ground-state  band of $^{168}$Er being the irrep with the next to lowest 
$E_{N_1N_2N_3}$ energy given for $^{168}$Er in Table \ref{eq:tab.1}.  But they are inconsistent with the much less deformed irrep with 
$N_0(\lambda_0\mu_0) = 814(30\;8)$ and a slightly lower value of
 $E_{N_1N_2N_3}$, which is the most deformed irrep that could be obtained in the standard spherical shell model.  What is especially notable is that 
 \emph{to obtain an Sp(3,\,R) irrep,  with the large experimentally-observed deformation, it is necessary to go to  unperturbed spherical-harmonic-oscillator shell-model states of  $12 \hbar\omega$ above the lowest-energy irrep of 
 $N_0 = 814$.}

\subsection{The wave functions of a symplectic model calculation}

Several algebraic symplectic model calculations have been made, where by algebraic we  mean calculations with phenomenological Hamiltonians that are low-order polynomials in the Sp$(3,\Rb)$ Lie algebra.  Such calculations, which explore the physical content of the symplectic model and its idiosynchracies in preparation for more detailed microscopic studies, became relatively straightforward following the development of vector coherent state methods \cite{Rowe84,RoweRC84,RoweRG85} for calculating the matrix elements of the Sp(3,\,R) Lie algebra. 

\begin{figure}[htb]
\centerline{\includegraphics[width=4.5 in]{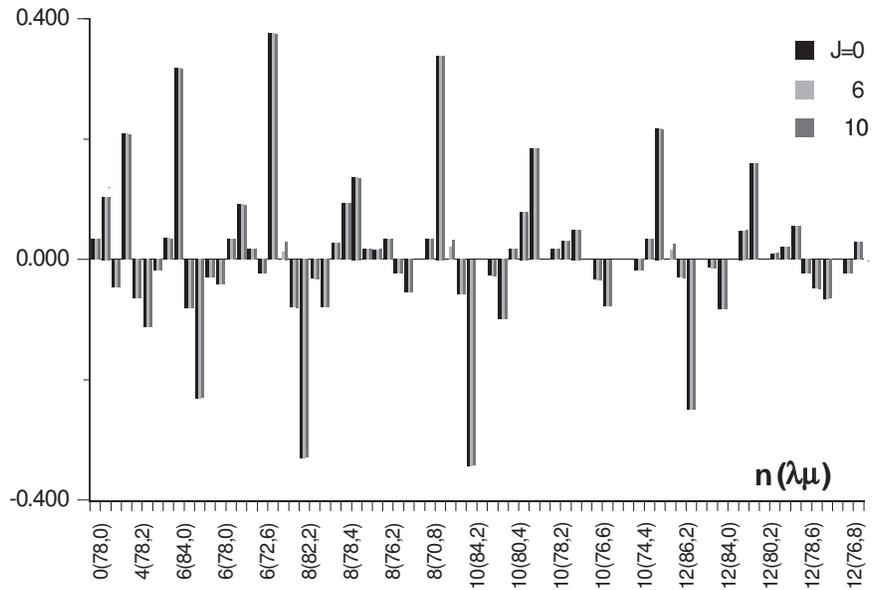}}
\caption{ Amplitudes for the wave functions of the 
$J=0$, 6, and 10, Sp(3.R) model states of figure \ref{fig.Sp3R166Er} in a U(3) basis.  The distribution of basis states includes states from 6 major harmonic-oscillator shells of positive parity that span a spherical harmonic-oscillator energy range of $12 \hbar\omega$.
Note that the SU(3) amplitudes of all states are  essentially independent of their angular momenta $J$.
\label{fig:166Erwfns}}
\end{figure}

Results of a  calculation  \cite{BahriR00} of the energy levels of the ground-state rotational band of $^{166}$Er  are shown in Fig.\ \ref{fig.Sp3R166Er}. The corresponding wave functions, shown in Fig.\ \ref{fig:166Erwfns}, are particularly revealing.
They are very different from those calculated both with algebraic and microscopic interactions for light nuclei%
\footnote{See the several results reviewed in Ref.\ \cite{Rowe85}.}
which tend to have their dominant components in the lowest spherical harmonic-oscillator shells and  contributions from higher shells that fall off rapidly over the space of a few shells.  In contrast, the amplitude of the lowest-grade U(3) irrep, $826.5(78\; 0)$, labelled $0(78\;0)$ in the figure, which is  the left-most amplitude shown in the figure, is seen to represent a minuscule component of the  states of the rotational band calculated.
Given that the lowest-grade U(3) irrep $826.5(78\; 0)$ is already at an excitation of $~12\hbar\omega$ relative to a spherical harmonic oscillator shell-model Hamiltonian, the amplitudes shown in this figure are a clear indication of the irrelevance of  shell-model calculations carried out in conventional shell-model spaces.
However, from a consideration of  Nilsson model   states for rotational bands  in highly-deformed nuclei, this  comes as no surprise.

\subsection{Some conclusions}

The above results lead to the following conclusions:
\begin{enumerate}

\item[(i)]  \emph{The partial ordering of Sp(3,\,R) irreps  by the spherical harmonic oscillator energies of their lowest-weight states, $N_0\hbar\omega$, is  inappropriate for deformed nuclei.}\smallskip

Recall that spin-orbit interactions also brings about a significant re-arrangment of the single-particle energy levels of the  spherical harmonic oscillator and that pseudo-SU(3) \cite{RajuDH73} and pseudo-symplectic models \cite{TroltenierBD95} have been proposed to take this into account. 
The above considerations suggest that the re-ordering of the relevant shell-model configurations due to deformation is an even larger concern  in strongly-deformed nuclei.  Even so, the mixing of symplectic irreps by spin-orbit and other interactions, notably pairing interactions, has to be  considered and is addressed briefly in the following section.  Thus, the potentiality for combining the pseudo-symplectic techniques and the above methods  is worth considering.

\item[(ii)] \emph{The symplectic model in combination with self-consistent mean-field considerations demonstrates the importance of the quadrupole component of the mean field and leads to the expectation of coexisting states of different deformation already in the low-energy domain.}\smallskip

  The best known examples of this in light nuclei are  the Hoyle state at 7.65 MeV in $^{12}$C and the first excited, 4-particle--4 hole, state in $^{16}$O,  Many other examples of shape coexistence are given in the recent review  of Heyde and Wood \cite{HeydeW11}.

\item[(iii)]  \emph{The dominance of highly deformed states in large open-shell nuclei, such as those of the rare-earth region, is an indication that, while the self-consistency formula, which orders Sp(3,\,R) irreps by  $E_{N_1N_2N_3} $,  provides  a first estimate of  the lowering of deformed states in heavy nuclei, it is probably an underestimate.}
\smallskip

It would appear to be an observed fact that highly-deformed  irreps fall  below the lowest-energy states of the spherical shell model  in large doubly-open-shell nuclei.
However, it unclear how far below they fall.
   This suggests the importance of seeking evidence  of spherical excited states in otherwise deformed nuclei, which might correspond to closed sub-shell configurations
(see \cite{HeydeW11}). It also suggests the importance of obtaining more reliable results from mean-field calculations in the spaces of Sp(3,\,R) irreps with meaningful nucleon-nucleon interactions. 

\end{enumerate}

\section{Microscopic calculations in Sp(3,\,R) bases}

The concern of this section is with the use of the subgroup chain
\be {\rm Sp(3,\,R)}  \supset {\rm U(3)} \supset {\rm SO(3)},\ee
 to define shell-model coupling schemes for  microscopic calculations.
 For a complete scheme, the basis states of this chain are combined with complementary spin-isospin states classified by the supermultiplet chain
\be {\rm U(4)} \supset  {\rm SO(4)} \supset {\rm SU(2)}_S \times {\rm SU(2)}_T
\ee
in an isospin formalism, or their neutron-proton counterparts,

Enormous progress has  been made by Dytrych et al.\ 
\cite{DytrychSBDV08,DytrychSDBV08}  in the development of efficient shell-model calculations in Sp(3,\,R) $\supset$ SU(3) bases and more general multi-shell SU(3) bases for no-core shell model calculations.
These developments are of fundamental importance for showing that the many-nucleon theory of nuclear physics can be derived with nucleon interactions    obtained from experiment, meson theory, and effective field theory (see lectures of Machleidt).

It is to be hoped that many of the essential techniques developed in 
\cite{DytrychSBDV08,DytrychSDBV08} will also apply to heavy nuclei.
The problem is the explosion of the dimensions of the no-core shell model approach with the number of spherical harmonic oscillator shells in the active space of the calculation, which Fig.\ \ref{fig:166Erwfns}, for example, shows can be be very large for heavy deformed nuclei.  Moreover, as this overview indicates, they are expected to lie well above the lowest-energy harmonic-oscillator shells.
Thus, we explore a complementary approach for applications to well-deformed states in rotational nuclei which reduces the dimensions of  the Sp(3,\,R) spaces needed for converged results by several orders of magnitude.

\subsection{Calculations within a single Sp(3,\,R) irrep}

We consider a generator coordinate  approach based on a procedure proposed by Filippov, Okhrimenko,  Vasilevsky, Vassanji, and others
 \cite{FilippovO80,VasilevskySF80,VassanjiR82,VassanjiR84} and further developed by Carvalho, and Rowe
 \cite{CarvalhoVR93,CarvalhoR97,CarvalhoRKB02}.

This approach is founded on the observation that the real submanifold ${\cal R}$ of Sp(3,\,R) coherent states (\ref{eq:realSp3Rcohstates}) spans the Hilbert space of the corresponding Sp(3,\,R) irrep.  Thus, a basis for a truncated subspace is given by a discrete set of states
\be |N_0(\lambda_0\mu_0) ijk\rangle = 
\hat R(\Omega_i) \hat U({\bf d}_j) \hat R(\Omega'_k) 
|N_0(\lambda_0\mu_0) \rangle,
\ee
where $|N_0(\lambda_0\mu_0) \rangle$ is the lowest-weight state for the irrep.
The notable fact, shown in Ref.\ \cite{CarvalhoRKB02}, is that converged results  are obtained using this method, to a given level of accuracy, with very few points when  chosen in an optimal way.

Good choices can be made following a computation of the mean field-energy
$E(\Omega {\rm d}\Omega')$ defined by Eq.\ (\ref{eq:E(OmdOm'}).
For a rotationally invariant  Hamiltonian, this energy is independent of 
$\Omega$, i.e., $E(\Omega {\rm d}\Omega')=E({\rm d}\Omega')$.  Taking linear combinations of states of many $\Omega_i$ then amounts to angular-momentum projection and the generation of states with good angular-momentum quantum numbers for which techniques have been developed, for example, by Cusson and Lee \cite{CussonLee73}.
With only a single diagonal matrix $\bar{\bf d}$, namely that for which
$E(\bar{\bf d})$ is a minimum and only the null rotation for $\Omega'\in$ SO(3), it is found that rather good results are already obtained.  However, the results, for a fixed value of ${\bf d}$ do not allow for centrifugal stretching effects corresponding to changes in the mean values of the elements of ${\bf d}$ with increasing angular momentum.  Such effects are included by allowing ${\bf d}$ to vary with the angular momentum or by taking a  small fixed set of $\{{\bf d}_j\}$ about the mean-field minimum value $\bar{\bf d}$.  (The preponderance of experimental evidence, based on yrast B(E2) values, suggests nearly constant $d$ values.)
Optimal  
$\{{\bf d}_j\}$ sets can be determined along the lines considered by Carvalho \emph{et al.} \cite{CarvalhoRKB02} in consideration of the energy functions $E({\bf d})$ and the range of values of ${\bf d}$ about the optimal value 
$\bar{\bf d}$  for which
$\langle N_0(\lambda_0 \mu_0) | U^\dag({\bf d}) U(\bar{\bf d})|N_0 (\lambda_0\mu_0)\rangle$ is non-negligible.
Similar choices can be made for $\{\Omega'_k\}$.

Applications of this approach to date obtain  promising results with just a few 
$\{{\bf d}_j\}$ and neglect of the   $\Omega'$ degree of freedom. 
Neglect of the $\Omega'$ degree of freedom has an interesting physical significance.
Note that  an $N_0(0\;0)$ irrep, in which $\Omega'$ does not contribute,
contains the spectrum of U(3) irreps
\be \big\{ \{N_0 +n (n_1-n_2, n_2-n_3)\}, \; n = 0, 2, 4, \dots \big\} , \label{eq:BMY3irreps}\ee
where $n_1$, $n_2$, and $n_3$ are even integers and 
$n_1\geq n_2 \geq n_3 \geq 0$.
Thus, as illustrated in Fig,\ \ref{fig:SU3irreps},
the U(3) spectrum of states for a generic Sp(3,\,R) 
$\langle N_0(\lambda_0\mu_0)\rangle$ is given by the tensor products of the
SU(3) irrep $(\lambda_0  \mu_0)$ with the U(3) irreps in the set 
(\ref{eq:BMY3irreps}).  With suppression of the $\Omega'$ degree of freedom, some of the irreps in this tensor product will be missing.  However, no irrep of the stretched product form
\be \{N_0 +n (\lambda_0+n_1-n_2, \mu_0 + n_2-n_3)\} , \quad n = 0, 2, 4, \dots
\ee
is omitted.  But, which precisely which irreps are missing is not obvious to me at this time.  This needs investigation and, if necessary, the $\Omega'$ degree of freedom can be included.

The technology for the efficient application of the above generator coordinate procedure remains to be fully developed.
So far, it has been applied  to Sp$(3,\Rb)$ irreps of maximal space symmetry; these are spaces of zero neutron and proton intrinsic spins.
More generally, it is necessary to consider  an irrep
 of the direct-product group ${\rm Sp(3,\,R)} \times {\rm SU(4)}$, in an isospin formalism, or the tensor product of two irreps of the direct product
 ${\rm Sp(3,\,R)} \times {\rm SU(2)}$, in a neutron-proton formalism.

\subsection{Mixed-representation calculations in an 
Sp(3,\,R) $\mathbf\supset$ U(3) basis and quasi-dynamical symmetry}

The development of efficient programs  for the calculation of states of even-even nuclei within spaces of multiple Sp(3,\,R) irreps of maximum space symmetry (i.e., spaces of zero neutron and proton intrinsic spin) will undoubtedly be achieved  in the near future.
Exploratory calculations within such spaces are needed, starting with model interactions, to identify the nature of any problems that might arise, before the bigger task of elaborating apposite techniques for mixed representations with non-zero intrinsic spins and  microscopic interactions is tackled.
This task would appear to be challenging but not insurmountable.
Given that  a relatively small number of optimally-chosen generator-coordinate basis states is needed, for each Sp$(3,\Rb)$ irrep, to obtain an acceptable level of converged solutions, and because rotational structure is much more clearly defined in heavy nuciei,
 it is possible that microscopic calculations in heavy deformed nuclei will eventually prove to be little, if any, more difficult than for the less deformed states of light nuclei.  


Calculations in spaces of mixed Sp(3,\,R) irreps are necessary to understand the influences on rotational structures arising from various Sp(3,\,R) symmetry-breaking interactions, such as spin-orbit interactions and pairing interactions.
Some years ago there was concern \cite{Ravello95} that, while systems with a dynamical symmetry can be handled with relative ease, systems with two or more competing dynamical symmetries would generally be intractable.
However, it was also known that, if the dynamics associated with one of a competing pair of dynamical symmetries was slow and the other fast,  there could be an adiabatic separation of the variables along the lines proposed by Born and Oppenheimer \cite{BornO27}.
This concept was therefore incorporated into the language of group theory in terms of the precisely-defined mathematical concept of an embedded representation \cite{RoweRR88}. Such representations were subsequently observed to arise approximately in many situations of competing dynamical symmetries and were described as quasi-dynamical symmetries 
\cite{RoweCancun04}.

For a review of  systems with competing dynamical symmetries see 
\cite{CejnarJC10}.  Such systems commonly exhibit  a quantum phase transition  between a phase in which one dynamical symmetry is dominant and another phase in which a  competing dynamical symmetry is dominant.
Moreover, the transition between the two phases typically becomes sharper for a many-particle system as the particle number is increased.
It appears that such a dominant symmetry 
is almost invariably characteristic of a quasi-dynamical symmetry.

The prototype of a quasi-dynamical symmetry is given by a system that has the dynamical symmetry of a rigid rotor with irreps defined by fixed intrinsic values of its quadrupole moments.  When the  rigid-rotor irreps  are mixed by dynamical symmetry-breaking interactions, the system becomes a soft rotor  with a distribution of intrinsic quadrupole moments.  However, in the adiabatic limit, 
in which the centrifugal and Coriolis forces are weak,
the intrinsic distribution remains constant over a range of low-energy angular-momentum states of its ground-state rotational band.  In isolation, these states then have the same relative properties as those of a rigid-rotor; their non-rigid intrinsic structures are only revealed by generally small but non-zero transition matrix elements to states of excited bands.

\begin{figure}[tph]
\centerline{\includegraphics[width=2.1in]{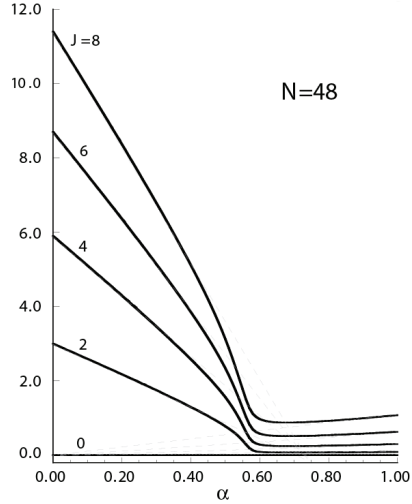}}
\caption{The low-energy spectrum for the model Hamiltonian (\ref{eq:Hsu2su3}) as a function of $\alpha$ for a system of $N=48$ particles.
\label{fig:su2su3}}
\end{figure}

The following example \cite{BahriRW98} of an $N=48$ particle system with a Hamiltonian, given as function of a control parameter $\alpha$ by
\be \hat H(\alpha) 
= \hat H_0 + (1-\alpha) \hat H_{\rm SU2} + \alpha \hat H_{\rm SU3} , 
\label{eq:Hsu2su3}\ee
illustrates a close approach to a sharp phase transition from one phase to  another as $\alpha$ is increased past a critical point.
The low-energy spectrum of this Hamiltonian is shown, as a function of $\alpha$ in Fig.\ \ref{fig:su2su3}.
When $\alpha = 0$ the Hamiltonian has an SU(2) $\subset$ USp(6)  dynamical symmetry and the system is in a superconducting vibrational phase.
And, when $\alpha = 1$ the Hamiltonian has an SU(3) $\subset$ USp(6)  dynamical symmetry and the system is in an adiabatic SU(3)-rotational phase (with low excitation energies).  For intermediate values of $\alpha$ the two dynamical symmetries are in competition.  The remarkable fact is that  the spectrum is characteristic of the SU(2) phase for $\alpha \lesssim 0.55$ and characteristic of the SU(3) phase for $\alpha \gtrsim 0.60$.  Moreover, as the particle number is increased the transition point becomes increasingly sharp.
%

\begin{figure}[pht]
\centerline{\includegraphics[width=3.8 in]{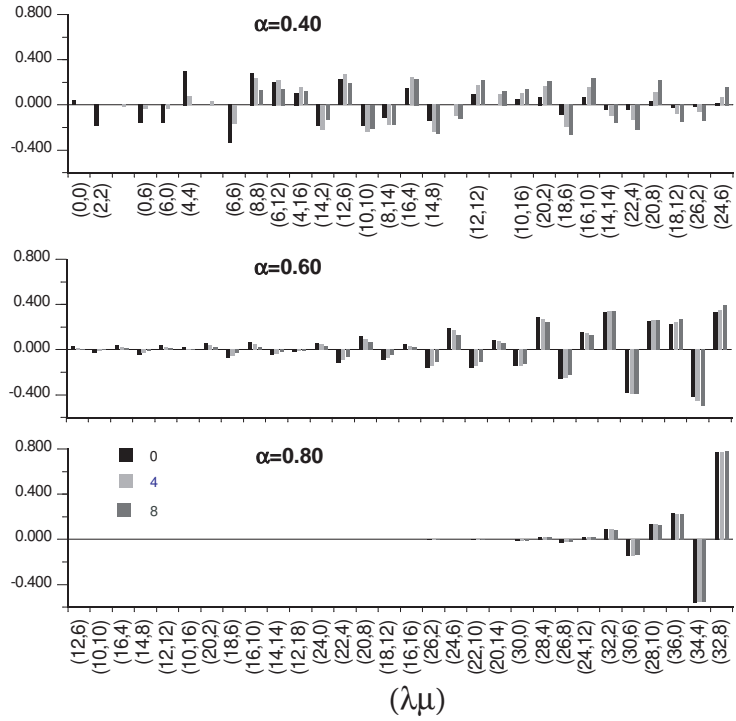}}
\caption{Histograms of the SU(3) amplitudes for the lowest energy $L=0$ to 8 eigenstates of the  Hamiltonian (\ref{eq:Hsu2su3}) for four values of $\alpha$.\label{fig:P+Qwfns}}
\end{figure}

The corresponding wave functions for three values of $\alpha$ are shown in Fig.\ 
\ref{fig:P+Qwfns}.
What is notable is the sudden change in the character of the wave functions as 
$\alpha$ increases past its critical value of $\approx 0.58$
As $\alpha$ increases above the critical point, the SU(3) amplitudes show a  reduced mixing of irreps but, more significantly, the mixed SU(3) structure of each state of the  quasi-SU(3) rotational band rapidly becomes the same for each of the states in the band.  It is as if all the SU(3) irreps were equivalent.  It appears that only SU(3) irreps with similar quantum numbers are mixed strongly by symmetry-breaking interactions.

The reduction in the mixing of basis states belonging to widely different irreps of a dominant dynamical symmetry is illustrated and  can be understood in the 
$^{12}$C and $^{16}$O calculations mentioned above.   For these nuclei, the mixing of the spherical harmonic oscillator lowest-energy configurations with those that form the highly deformed  first-excited states  is  much suppressed by the fact that all matrix elements of one- and two-body operators between the low-lying Sp(3,\,R) irreps of the very different deformations, shown in Table \ref{eq:tab.1}, are precisely zero.  These irreps can only mix indirectly  via higher-lying irreps.  For example, the lowest-grade states of the excited irreps of $^{12}$C and $^{16}$O require the excitation of four particles relative to the ground-state irrep.  Consequently, all matrix elements between states of the dominant irreps contributing to the ground and excited states are identically zero for a Hamiltonian with only two-body interactions.

\section{Summary and discussion}

I hope to have demonstrated in these talks that the well-developed  group-theoretical and algebraic methods relating to symmetry provide important insights and powerful tools for investigating nuclear structure.  In conjunction with geometry, analysis, and other standard tools, such as mean-field and coherent-state theory,  they provide fundamental tools for studying the many interesting phenomena displayed in nuclear physics.

A  message I want to emphasise is the importance of determining the domain of validity of a successful phenomenological model and of
pursuing its relationship with other more sophisticated models, e.g., the many-nucleon theory of the nucleus.  Investigating the limitations of a model is every bit as important as investigating its successes; it is probably more important.
And establishing its microscopic foundations is what enables the model to contribute to the deeper understanding of the system it models.

 I have focused on the emergence of the symplectic model,
exposed its roots in the Bohr collective model, and pursued its role in the development of a microscopic shell-model theory of rotational nuclei. However, I would not want to create the impression that the symplectic model stands alone.
The primary pillars of nuclear structure remain. They include: the shell model, the collective model, the unified models, the pairing model, and the Hartree-Fock and Hartree-Bogolyubov mean-field theories. The symplectic model builds on these foundations and, in parallel with the algebraic pairing model, it defines a shell-model coupling scheme that enables one to make use of what one learns, from model fits to data, in the design of a more complete microscopic  theory.
Like the pairing model, which supplies the standard $JJ$ coupling scheme of the spherical shell model and provides a shell-model theory of singly-closed shell nuclei, the symplectic model supplies a coupling scheme that relates to shell-model bases expressed both in terms of spherical harmonic oscillator and anisotropic harmonic oscillator wave functions.  Thus, as discussed in the text, it provides a shell-model theory of particular relevance to strongly deformed rotational states of nuclei.

Some possibly novel perspectives have been presented.  However, what is satisfying that the symplectic model embraces the views assembled from the vantage points of many developments and sets them in a new light.
For example, an extension of the spherical shell-model with spherical harmonic oscillator wave functions  to the Nilsson model \cite{Nilsson55} with anisotropic harmonic oscillator wave functions was made already in 1955. However, the extension was designed only for the purpose of describing the states of added nucleons interacting with a deformed rotational core nucleus.
Nevertheless, it was envisioned by many that the Nilsson model should somehow be extendable to a shell-model for deformed nuclei with non-spherical basis wave functions.
The symplectic model, which relates many-particle Nilsson model states to lowest-weight states of Sp(3,\,R) irreps, makes this vision a reality.
Moreover, as we have discussed, it leads to a practical generator coordinate approach to Sp(3,\,R) model calculations with microscopic interactions and approximations to mixed representation calculations with the inclusion of spin-orbit and other Sp(3,\,R) symmetry-breaking interactions.

The symplectic model gives the deformed shell model a fundamental foundation.
In particular, it resolves  concerns about completeness and orthogonality.   In the first place,  states associated with different Sp(3,\,R) irreps and different angular momenta and spin are strictly orthogonal.  
Moreover, it is known that a geometric space of Sp(3,\,R) coherent states spans the Hilbert space for an Sp(3,\,R) irrep.  Thus, the states of each such space are orthogonal to all other such spaces and, together, they span the whole  many-nucleon Hilbert space.
Thus, it is only necessary to orthogonalise basis states with common values for the associated quantum numbers.  This is an enormous advantage over similar projected Hartree-Fock methods both computationally and conceptually; e.g., it retains a knowledge of what shell model spaces and Sp(3,\,R) irreps are  included in a calculation.

In conclusion, it should be emphasised that the construction of a shell-model theory of doubly open-shell nuclei is far from a minor topic in nuclear structure physics.  Certainly  doubly closed- and singly closed-shell nuclei have contributed a lot to the understanding of nuclear structure.   However, such nuclei represent a small fraction of the nuclei available for study.  Moreover, it is rapidly becoming apparent \cite{HeydeW11} that strongly deformed states, generally associated with doubly-open shell nuclei, are common among the excited states of singly and doubly closed-shell nuclei in which both the neutron and proton degrees become active,

\bibliographystyle{aipproc}   

\bibliography{master}

\IfFileExists{\jobname.bbl}{}
 {\typeout{}
  \typeout{******************************************}
  \typeout{** Please run "bibtex \jobname" to optain}
  \typeout{** the bibliography and then re-run LaTeX}
  \typeout{** twice to fix the references!}
  \typeout{******************************************}
  \typeout{}
 }

\end{document}